%% file: main.tex
\documentclass[10pt,twocolumn,letterpaper]{article}

\usepackage{iccv}              


\definecolor{iccvblue}{rgb}{0.21,0.49,0.74}
\usepackage[pagebackref,breaklinks,colorlinks,allcolors=iccvblue]{hyperref}
\usepackage{tabularx}
\usepackage{float}
\usepackage{bm}
\usepackage{soul}
\usepackage{xcolor}
\usepackage{multirow}
\usepackage{booktabs}
\usepackage{subcaption}
\usepackage{graphicx}
\usepackage[accsupp]{axessibility}  
\def\paperID{7082} 
\def\confName{ICCV}
\def\confYear{2025}

\title{OccluGaussian: Occlusion-Aware Gaussian Splatting \\ for Large Scene Reconstruction and Rendering}

\author{
	Shiyong Liu\textsuperscript{1$\ast$}\;\;\;
	Xiao Tang\textsuperscript{1$\ast$}\;\;\;
	Zhihao Li\textsuperscript{1$\ast$}\;\;\;
	Yingfan He\textsuperscript{2}\\
	Chongjie Ye\textsuperscript{2}\;\;\;
	Jianzhuang Liu\textsuperscript{3}\;\;\;
        Binxiao Huang\textsuperscript{4}\;\;\;
	Shunbo Zhou\textsuperscript{5}\;\;\;
	Xiaofei Wu\textsuperscript{1\textdagger}\\
	{\textsuperscript{1}Huawei Noah's Ark Lab} \;\;\;\; {\textsuperscript{2}} The Chinese University of Hong Kong (Shenzhen)\;\;\;\; \\{\textsuperscript{3}Shenzhen Institutes of Advanced Technology}\;\;\;\; {\textsuperscript{4}The University of Hong Kong}\;\;\;\; \\{\textsuperscript{5}Huawei Embodied Intelligence Lab}\\
}

\begin{document}
\twocolumn[{%
	\renewcommand
	\twocolumn[1][]{#1}%
	\maketitle
	\begin{center}
		\centering
		 \vspace{-15pt}
		\includegraphics[width=1\textwidth]{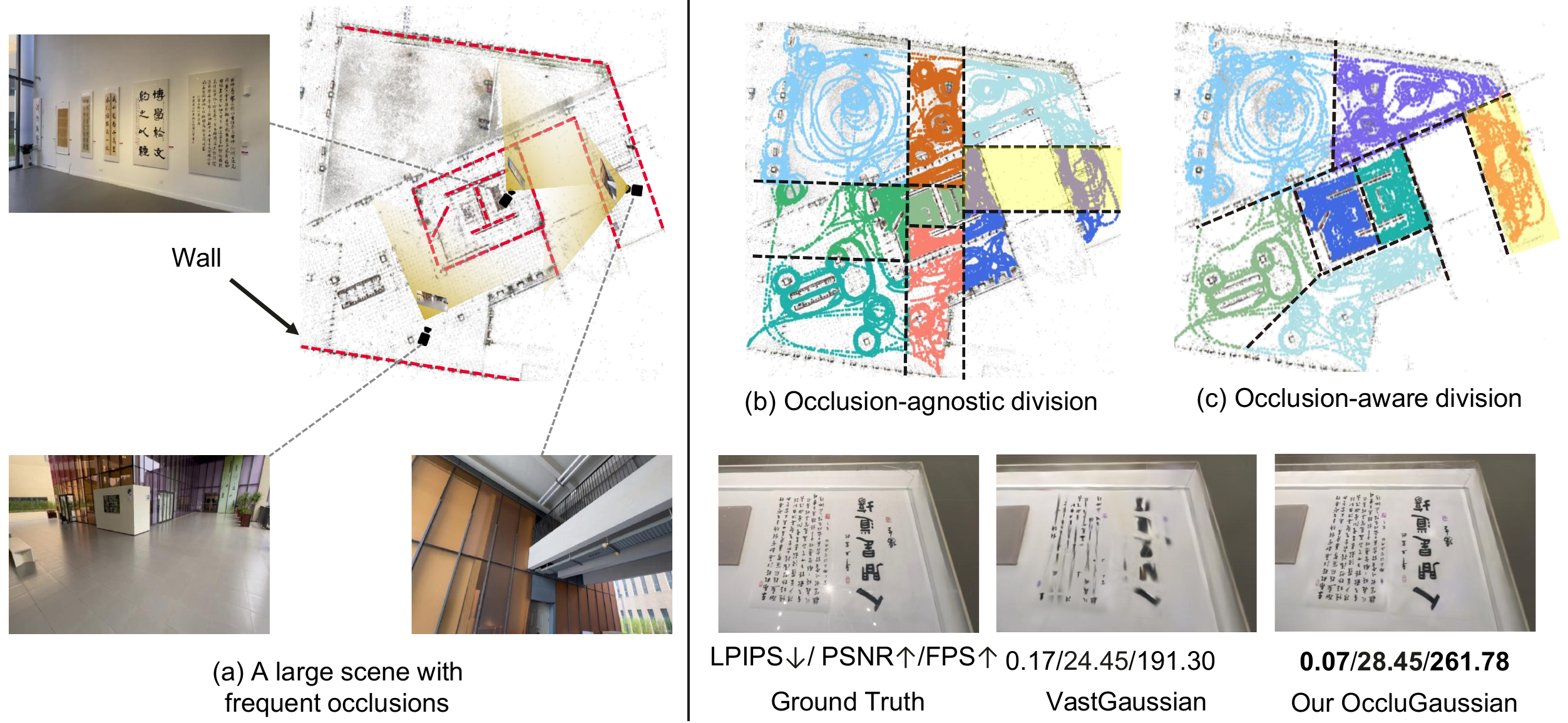}
		\captionof{figure}{(a) In large scene reconstruction from ground-level captures, there are frequent occlusions such as walls and buildings. (b) Previous scene division methods are occlusion-agnostic, so they produce regions with severe internal occlusions, leading to poor reconstruction results. (c) We propose an occlusion-aware division method to generate regions that better align with the scene layout (see the two regions highlighted in yellow in (b) and (c)), thus improving the reconstruction quality significantly. Besides, we introduce a region-based rendering technique to accelerate the rendering speed of 3D Gaussian splatting in a large scene with massive primitives.}
		\label{fig:teaser}
	\end{center}
}]
\renewcommand{\thefootnote}{}
\footnotetext{\small $^{\ast}$ Equal contribution \quad $^{\dag}$ Corresponding author}
\setcounter{footnote}{0}
\input{sec/0_abstract}    
\input{sec/1_intro}
\input{sec/2_related_work}
\input{sec/3_method}
\input{sec/4_experiments}
\input{sec/5_conclusion}
{
\newpage
    \small
    \bibliographystyle{ieeenat_fullname}
    \bibliography{main}
}

\newpage
\input{sec/X_suppl}


\end{document}

%% file: sec/0_abstract.tex
\begin{abstract}
In large-scale scene reconstruction using 3D Gaussian splatting, it is common to partition the scene into multiple smaller regions and reconstruct them individually. However, existing division methods are occlusion-agnostic, meaning that each region may contain areas with severe occlusions. As a result, the cameras within those regions are less correlated, leading to a low average contribution to the overall reconstruction. In this paper, we propose an occlusion-aware scene division strategy that clusters training cameras based on their positions and co-visibilities to acquire multiple regions. Cameras in such regions exhibit stronger correlations and a higher average contribution, facilitating high-quality scene reconstruction. We further propose a region-based rendering technique to accelerate large scene rendering, which culls Gaussians invisible to the region where the viewpoint is located. Such a technique significantly speeds up the rendering without compromising quality. Extensive experiments on multiple large scenes show that our method achieves superior reconstruction results with faster rendering speed compared to existing state-of-the-art approaches. Project page: \url{https://occlugaussian.github.io}.
\end{abstract}
\vspace{-2ex}

%% file: sec/1_intro.tex
\section{Introduction}
Large scene reconstruction needs to process a huge amount of geometry and appearance information in 2D images to recover 3D structures for novel view synthesis.
%
%
It is crucial in various applications, such as autonomous navigation~\cite{dalal2024gaussian, ann2016study, thrun2002robotic, tancik2022block, he2024neural}, cultural heritage preservation~\cite{scopigno2017digital}, and immersive virtual/augmented reality~\cite{mortara2014learning, fei20243d, dalal2024gaussian}, \etc.

Recent works in large scene reconstruction~\cite{turki2022mega, tancik2022block, xu2023grid, liu2024citygaussian, lin2024vastgaussian, kerbl2024hierarchical} mostly take radiance fields as the basic 3D representation, \eg, Neural Radiance Fields (NeRF)~\cite{mildenhall2021nerf} and 3D Gaussian Splatting (3DGS)~\cite{kerbl20233d}. 
{NeRF-based approaches struggle to scale for large scenes with rich details, as their implicit representations demand substantial resources for both training and rendering. 3DGS, a primitive-based rasterization technique, also faces scalability issues due to its memory-intensive representation, which quickly exceeds the capacity of high-end GPUs.}
{To address these issues, a divide-and-conquer strategy is commonly applied}, which partitions the scene into several smaller and more manageable regions. These regions are then individually reconstructed and finally merged to a complete model.
Existing scene division strategies are mainly based on camera positions or point clouds~\cite{lin2024vastgaussian, liu2024citygaussian, tancik2022block, xu2023grid, kerbl2024hierarchical}. While these approaches are suitable for occlusion-free scenes like aerial imagery or open spaces, they overlook scene layouts and occlusions, which are commonly present in indoor environments, as shown in Fig.~\ref{fig:teaser}.
{Comparing the two regions yellow highlighted in Fig.~\ref{fig:teaser}(b) and Fig.~\ref{fig:teaser}(c)}, the former have fewer training cameras with visible parts in common. The cameras for one area occupy a certain amount of training time and resources, but contribute little to the reconstruction of the other. In other words, putting them together involves too many training cameras whose average contribution to the region's reconstruction is thus significantly reduced.
%

%
%
%

In this paper, we introduce Occlusion-aware 3D Gaussian splatting (OccluGaussian) for large scene reconstruction and rendering. OccluGaussian incorporates an occlusion-aware scene division strategy that considers scene layout and occlusions. {It enhances training efficiency and resource allocation by focusing on cameras that substantially contribute to the reconstruction of specific areas, rather than diluting efforts across less relevant regions,} {thus finally improving the reconstruction quality.}
Specifically, we {first} create an attributed view graph where nodes represent cameras with position features and edges denote their visibility correlations, {based on the co-visibilities of the capturing cameras.}
{Next}, we apply a graph clustering algorithm to this graph, yielding division results that align with the scene layout, as shown in Fig.~\ref{fig:teaser}(c).
Training cameras in such regions have stronger correlations, enhancing their average contribution and leading to improved reconstruction results. 
Finally, we reconstruct these regions individually and merge them together to produce the complete model.


{The reconstructed complete model often contains a huge amount of Gaussians, leading to slow rendering speeds. Existing methods~\cite{liu2024citygaussian, kerbl2024hierarchical} often apply the level-of-details (LoD) technique to dynamically prune unnecessary small and distant Gaussians to boost the rendering efficiency.
However, these methods are still occlusion-agnostic and process all 3D Gaussians within the view frustum of the rendering camera. The occluded Gaussians that are invisible from the camera can be culled in advance to accelerate rendering without any quality drop.
To this end, we further propose a region-based rendering technique to accelerate the rendering of the large scene.}
Specifically,
{for each region generated from our scene division strategy,}
we record the visible 3D Gaussians of all training cameras located in it. 
During rendering, we process only the recorded 3D Gaussians of the region in which the rendering camera is located.
{Since our scene division strategy is occlusion-aware, we can effectively prune the occluded Gaussians that are invisible from the camera.}
This technique achieves much faster rendering without transition artifacts and noticeable loss in quality.

Our contributions are summarized as follows:
\begin{itemize}
\item We propose an occlusion-aware scene division strategy that considers the scene layout and camera co-visibilities. The resulting regions barely contain occlusions, and the corresponding training cameras have a higher average contribution, leading to improved reconstruction results.
\item We present a region-based rendering technique that accelerates 3D Gaussian splatting in large scenes. It eliminates much of the time-consuming processing of invisible 3D Gaussians, boosting rendering speeds without noticeable quality degradation.
\item We conduct extensive experiments on several large-scene datasets, and demonstrate that OccluGaussian achieves superior rendering quality and faster rendering speed compared to previous state-of-the-art methods.
\end{itemize}

%% file: sec/2_related_work.tex
\section{Related Work}

\subsection{Large Scene Reconstruction}
3D reconstruction of large scenes from captured images has been a popular research topic for decades.
Traditional methods use structure-from-motion (SfM) algorithms to generate sparse point clouds of the scene or further extract dense point clouds and meshes via multiview stereo methods~\cite{fruh2004automated, snavely2006photo,li2008modeling,pollefeys2008detailed,agarwal2011building,schonberger2016structure,zhu2018very,furukawa2010towards,goesele2007multi}.
Recently, Neural Radiance Fields (NeRF)~\cite{mildenhall2021nerf,xu2023grid,zhenxing2022switch,tancik2022block} and 3D Gaussian Splatting (3DGS)~\cite{kerbl20233d,lin2024vastgaussian,liu2024citygaussian,chen2025dogs,zhao2024scaling,wang2024pygs, lin2025decoupling, liu2024mirrorgaussian} have been widely applied to large-scale scene reconstruction, as they outperform point clouds and meshes for novel view synthesis.
A divide-and-conquer strategy is commonly applied for both NeRF and 3DGS methods.
Such as, BlockNeRF~\cite{tancik2022block}, as a pioneering work, explicitly splits city-scale scenes into multiple blocks, then optimizes individual NeRF model for each, and seamlessly combining them by aligning their appearances.
Mega-NeRF~\cite{turki2022mega} also uses a grid-based division, and assigns each pixel's ray to different grids that the ray intersects as it passes through the scene when optimizing NeRF models.
NeRF-XL~\cite{li2024nerf} distributes the divided blocks across multiple GPUs, allowing to reconstruct arbitrarily large scenes with more parameters and faster speed.
As for 3DGS-based methods, VastGaussian~\cite{lin2024vastgaussian} uniformly divides the scene and introduces a progressive partitioning strategy to ensure sufficient supervision for each block.
CityGaussian~\cite{liu2024citygaussian} splits a large scene in the contracted space to balance the optimization workload for each block.
Distributed training systems with multiple GPUs~\cite{chen2025dogs,zhao2024scaling} are also explored to achieve faster optimization speed or hold more Gaussian primitives to improve reconstruction quality.
However, these division approaches are mainly based on camera positions, and overlook the visibility correlations between cameras,
making them occlusion-agnostic. While such methods work well for aerial captures in popular large-scene datasets, they are less suitable for ground-level captures where occlusions occur frequently.

\subsection{Camera Clustering}
In camera pose estimation of large-scale scenes, partitioning methods are commonly utilized to cluster cameras and segment scenes for parallel processing, which helps to reduce the computational complexity~\cite{karypis1997metis, sattler2012improving, li2010location, zhang2017distributed, ni2007out}. Some of them~\cite{ni2007out, zhang2017distributed, bhowmick2015divide, zhu2018very, chen2020graph} use graph cut algorithms to aggregate cameras acoording to their co-visibilities, e.g., Out-of-Core-BA~\cite{ni2007out} and COLMAP~\cite{schonberger2016structure} use the Metis graph partitioner~\cite{karypis1997metis}. 
Others~\cite{li2010location, sattler2012improving} use clustering algorithms based on the features of images and 3D key points. However, these methods often adopt only image or matching features, which are not sufficient enough to avoid occlusions within the segmented regions, as shown in Fig.~\ref{fig:clutering_methods_comparison}. In this paper, we propose a scene division technique that exploits both camera position and co-visibility information, which is occlusion-aware and capable of aligning with the scene layout to conduct partitioning.

\begin{figure*}[t]
  \centering
  \includegraphics[width=0.8\linewidth]{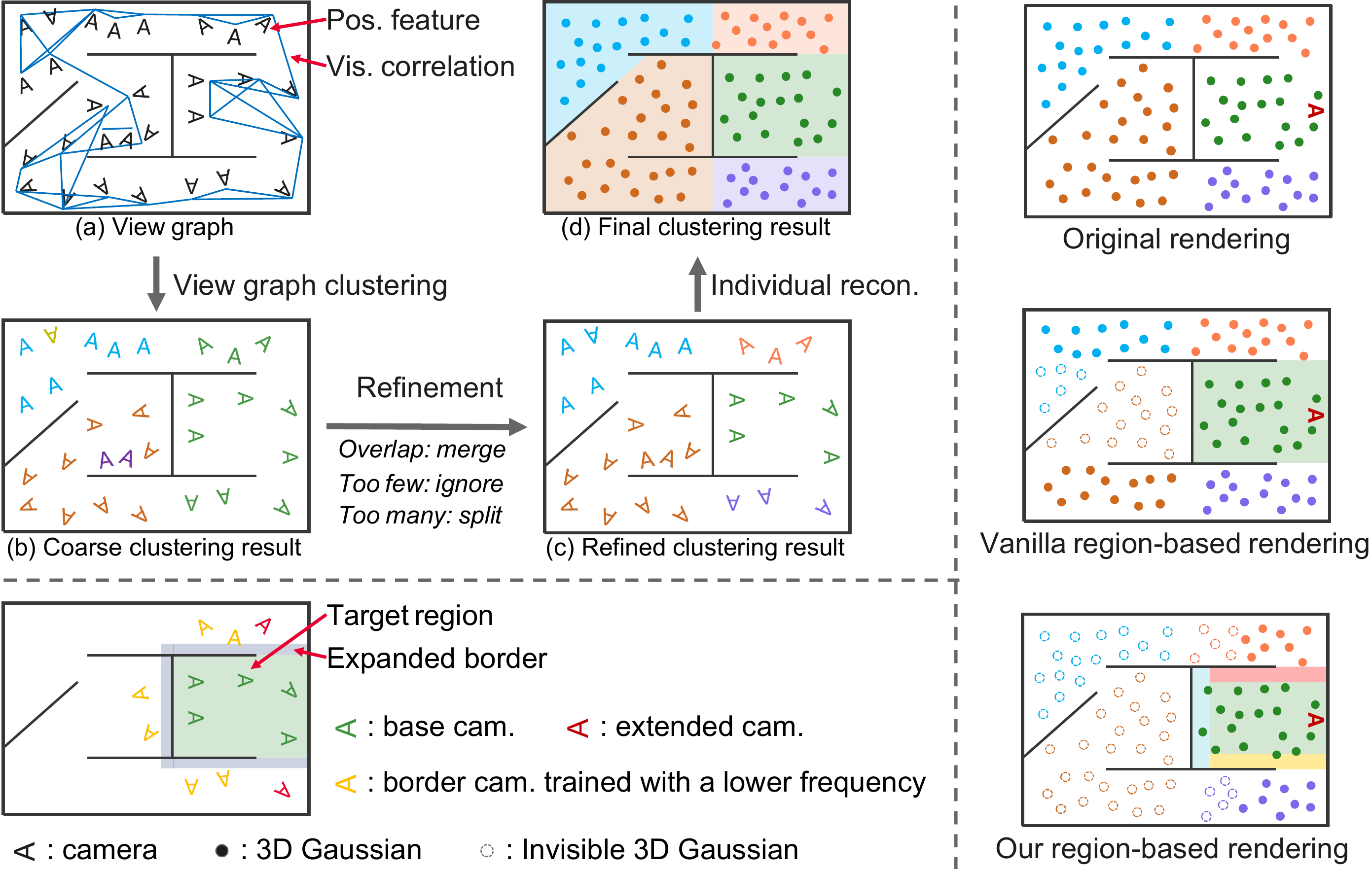}
  \caption{{\bf Overview of OccluGaussian.} 
  {\bf Top left:} To reconstruct a large scene, we divide it into multiple regions by adopting an occlusion-aware scene division strategy.
  (a) We first create an attributed view graph from the posed cameras, where nodes represent cameras with positional features, and edges represent visibility correlations between them.
  (b) A graph clustering algorithm is applied to the view graph to cluster the cameras into multiple regions, and (c) we further refine them to obtain more balanced sizes.
  (d) The region boundaries are calculated based on the clustered cameras. Each region is individually reconstructed and finally merged into a complete model. 
  {\bf Bottom left:} Each region is reconstructed using three sets of training cameras: base cameras located inside the region, extended cameras providing adequate visual content of the region, and border cameras used to constrain Gaussian primitives near the boundaries.
  {\bf Right:} We introduce a region-based rendering technique, which culls 3D Gaussians that are occluded from the region where the rendering viewpoint is located. Furthermore, we subdivide the scene into smaller sub-regions with fewer essential 3D Gaussians. This approach effectively reduces redundant computations and further boosts our rendering speed.
  }
  \vspace{-2ex}
  \label{fig:overview}
\end{figure*}

\subsection{3DGS Rendering Acceleration}
As large scenes contain a substantial number of Gaussian primitives, it is crucial to speed up 3DGS rendering for real-time performance.
Recently, several methods~\cite{lee2024compact, fan2024lightgaussian, navaneet2023compact3d, jiang2024hifi4g, girish2024eagles, morgenstern2024compact, niemeyer2024radsplat} have been proposed to compress 3DGS by reducing the number of Gaussian primitives, which also helps accelerate rendering in large scenes.
These methods prune 3D Gaussians that contribute minimally to the overall rendering, such as those with low opacities, to minimize the impact on visual quality.
Another popular approach for rendering acceleration customizes the rendering for large scenes by using different level-of-details (LoD) at varying distances.
For example, Octree-GS~\cite{ren2024octree} incorporates an octree structure to hierarchically organize 3D Gaussians into multiple levels.
CityGaussian~\cite{liu2024citygaussian} uses the compression algorithm from LightGaussian~\cite{fan2024lightgaussian} to generate different detail levels in a block-wise manner.
Hierarchical-3DGS~\cite{kerbl2024hierarchical} introduces a novel hierarchy for 3DGS, enabling efficient level-of-detail selection and interpolation.
From a different perspective compared to existing methods, OccluGaussian proposes a region-base rendering strategy, which culls occluded Gaussians that are invisible to the viewpoint in advance. This significantly reduces redundant computations and further boosts rendering speeds.


%% file: sec/3_method.tex
\section{Method}
We propose an occlusion-aware 3DGS framework (OccluGaussian) that achieves superior reconstruction quality and significantly faster rendering speed for large-scale scenes. 
The overview of OccluGaussian is given in~\cref{fig:overview}.
We first present a novel occlusion-aware scene division strategy based on attributed graph clustering in~\cref{sec:division}.
Next,~\cref{sec:reconstruction} details the optimization of each individual region and the merging process to obtain a complete model.
Finally, we introduce a region-based rendering technique to accelerate rendering in large scenes in~\cref{sec:rendering}.

\subsection{Occlusion-Aware Scene Division} \label{sec:division}
%
Previous division methods~\cite{turki2022mega, tancik2022block, xu2023grid, liu2024citygaussian, lin2024vastgaussian, kerbl2024hierarchical} typically ignore the scene layout and divide the scene uniformly based on camera positions or point clouds. A more effective division strategy should consider the occlusions in the scene, so that the cameras in each region have strong correlations and a higher average contribution to the reconstruction. Here, we introduce an occlusion-aware scene division strategy, as illustrated in the top-left part of ~\cref{fig:overview}.
%
%

\noindent{\bf Attributed view graph.}
First, we create a view graph of $n$ posed training cameras, which is an undirected attributed graph $\mathcal{G} = (\mathcal{V}, \mathcal{E}, X)$, where $\mathcal{V}$, $\mathcal{E}$ and $X$ denote the nodes, the edges with weights, and the feature matrix, respectively. 
Each node in $\mathcal{V} = \{v_1, v_2, \ldots, v_n\}$ corresponds to a camera.
To derive the edges $\mathcal{E}$, we directly utilize the structure-from-motion results obtained during the camera pose estimation.
An edge is established between two cameras if they share some visual contents in common, and we set its weight as the number of matched feature points between these two cameras. The set of edges $\mathcal{E}$ is finally represented as an adjacency matrix $A = [{a_{ij}}] \in \mathbb{R}^{n \times n}$.
%
%
%
The feature matrix $X = [\bm{x_1}, \bm{x_2}, \bm{\ldots}, \bm{x_n}]^\top \in \mathbb{R}^{n \times d}$ contains the 3D coordinates of each camera with positional encoding~\cite{mildenhall2021nerf}. 
%
Intuitively, occluded or distant cameras typically share minimal overlapped views, which can be distinguished in our occlusion-aware view graph.

\noindent{\bf View graph clustering.}
%
%
Then, we split the view graph $\mathcal{G}$ into multiple parts using graph clustering.
We employ a attributed graph clustering algorithm~\cite{zhang2023adaptive}, which performs graph convolution to generate smooth feature representations, followed by spectral clustering on the resulting features to group the nodes. The clustering process is outlined as follows:

Given the adjacency matrix $A$ and the degree matrix $D=\text{diag}(d_1, \cdots, d_n)$, ${d_i} =\sum\nolimits_{j=1}^n{a_{ij}}$, 
the symmetrically normalized graph Laplacian is defined as:
\begin{equation}\label{eq:laplacian}
    L_{s}=I-D^{-\frac12}AD^{-\frac12}.
\end{equation}
Graph convolution is defined as the multiplication of a graph signal $f$ with a graph filter $G$:
\begin{equation}\label{eq:graph convolution}
    \bm{\bar{f}}=G\cdot\bm f,  \quad
     G=(I-\frac12 L_s)^r, 
\end{equation}
where $\bm{\bar{f}}$ is the filtered graph signal, and $r$ is a positive integer. Each column of the feature matrix $X$ can be considered as a graph signal. By performing graph convolution on the feature matrix $X$, we obtain the filtered feature matrix $\bar{X}=GX$. 
Then, we calculate the similarity matrix $\bar{W}$:
\begin{equation}\label{eq:similarity_matrix}
\bar{W}=\frac12(\left|H\right|+\left|H^T\right|),  \quad {H} = \bar{X}\bar{X}^T. 
\end{equation}
{Since $\bar{W}$ indicates the closeness between nodes in $\mathcal{V}$, we apply spectral clustering algorithm~\cite{perona1998factorization} on $\bar{W}$ to cluster $\mathcal{V}$,} which correspond to the training cameras. Cameras that share a large overlapping view or are spatially close are clustered into the same region, as shown in~\cref{fig:overview}(b).

\noindent{\bf Determining the clustering number.}
Most clustering algorithms need a predefined number of clusters. To address this, we propose an adaptive approach to determine the optimal number of clusters. 
The process starts by selecting an initial cluster number $K$, enabling graph clustering to generate $K$ preliminary clusters. 
%
%
We then refine them by splitting clusters that contain too many cameras by further applying graph clustering, and by ignoring any cluster that either has too few cameras or whose convex hull is entirely covered by the convex hull of another cluster. 
These refinement steps are applied recursively until all clusters achieve a balanced number of cameras, thereby determining the optimal number of clusters for different scenes, 
(see~\cref{fig:overview}(c)). 

\noindent{\bf Boundary calculation.}
Finally, we calculate explicit boundaries of each region based on its clustered cameras. 
These boundaries are crucial for border expansion (\cref{sec:reconstruction}) and region-based rendering (\cref{sec:rendering}).
We use a linear classification method~\cite{cortes1995support} to derive decision functions as boundary lines.
Thanks to the occlusion-aware camera clustering, these boundary lines closely align with the scene layout.

\subsection{Individual Region Reconstruction} \label{sec:reconstruction}
After the occlusion-aware scene division, we reconstruct each region individually and finally merge them to form the complete model. In this subsection, we detail the strategy for selecting training cameras for each region and the algorithm used to seamlessly merge these regions.

\noindent{\bf Training camera selection.}
As pointed out in~\cite{lin2024vastgaussian,liu2024citygaussian}, it is crucial to select appropriate training cameras for 3D reconstruction.
Too few training cameras cannot provide sufficient supervision, while too many irrelevant cameras will waste computational resources and reduce the  average contribution of every camera.
%
We select three sets of training cameras as shown in the bottom-left part of~\cref{fig:overview}: 1) The base set, whose cameras are located within the region. 2) The extended set, whose cameras are outside the region but capture adequate visible content of it. We follow~\cite{liu2024citygaussian} to select the extended cameras based on their visibility contribution to the region.
3) The border set, whose cameras face the region but are occluded. These cameras help constrain Gaussian primitives near the boundaries for the final seamless merging. Without the border set, these Gaussian primitives can become very large or elongated, resulting in floaters or artifacts (see the qualitative results in the supplementary material).
Compared to existing methods, our occlusion-aware scene division strategy requires fewer extended cameras for adequate supervision to reconstruct the target region, increasing the average contribution per camera and thereby improving reconstruction quality.

%
%

\noindent{\bf Seamless region merging.}
We optimize 3D Gaussians for each region individually using its selected training cameras.
And then, we remove all Gaussian primitives from each region that lie outside the region to create sharp borders. Finally, these regions are merged to form a complete model.

\begin{figure*}[t]
  \centering	\includegraphics[width=0.95\linewidth]{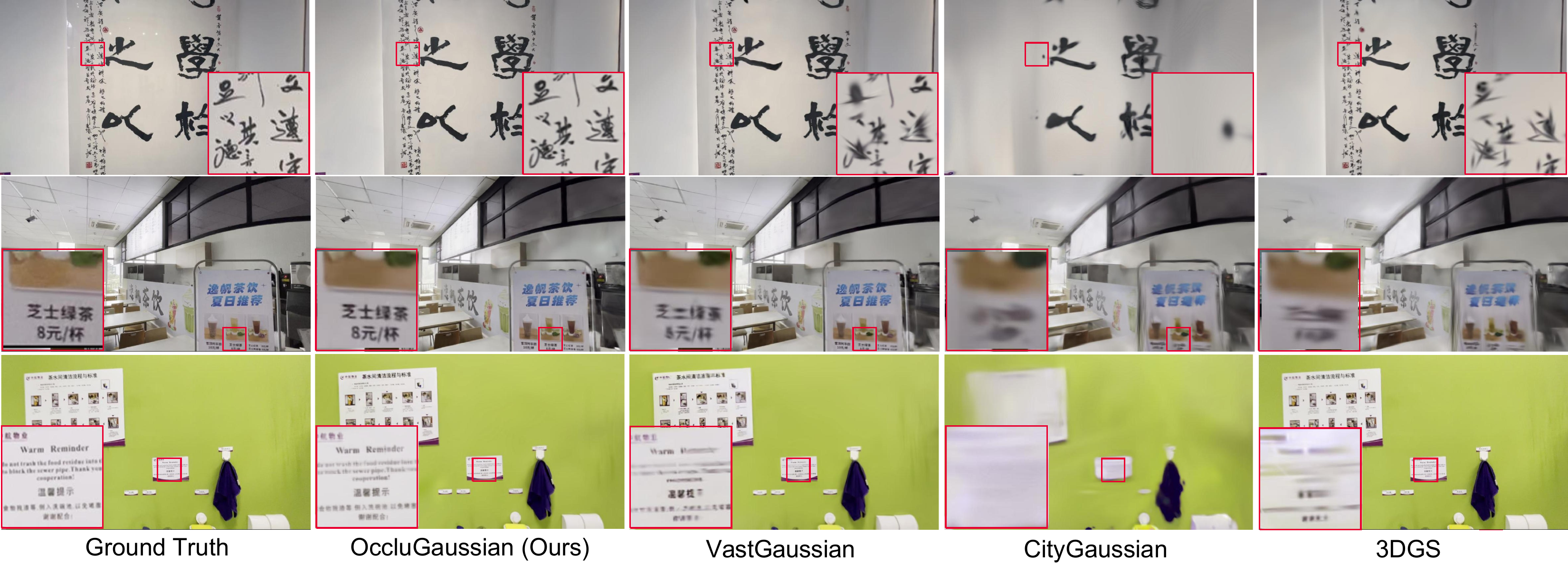}
	\caption{Qualitative comparison with SOTA methods on three large scenes in the OccluScene3D dataset.}
	\label{fig:visual_comparison}
\end{figure*}

\begin{figure*}[t]
  \centering	\includegraphics[width=0.95\linewidth]{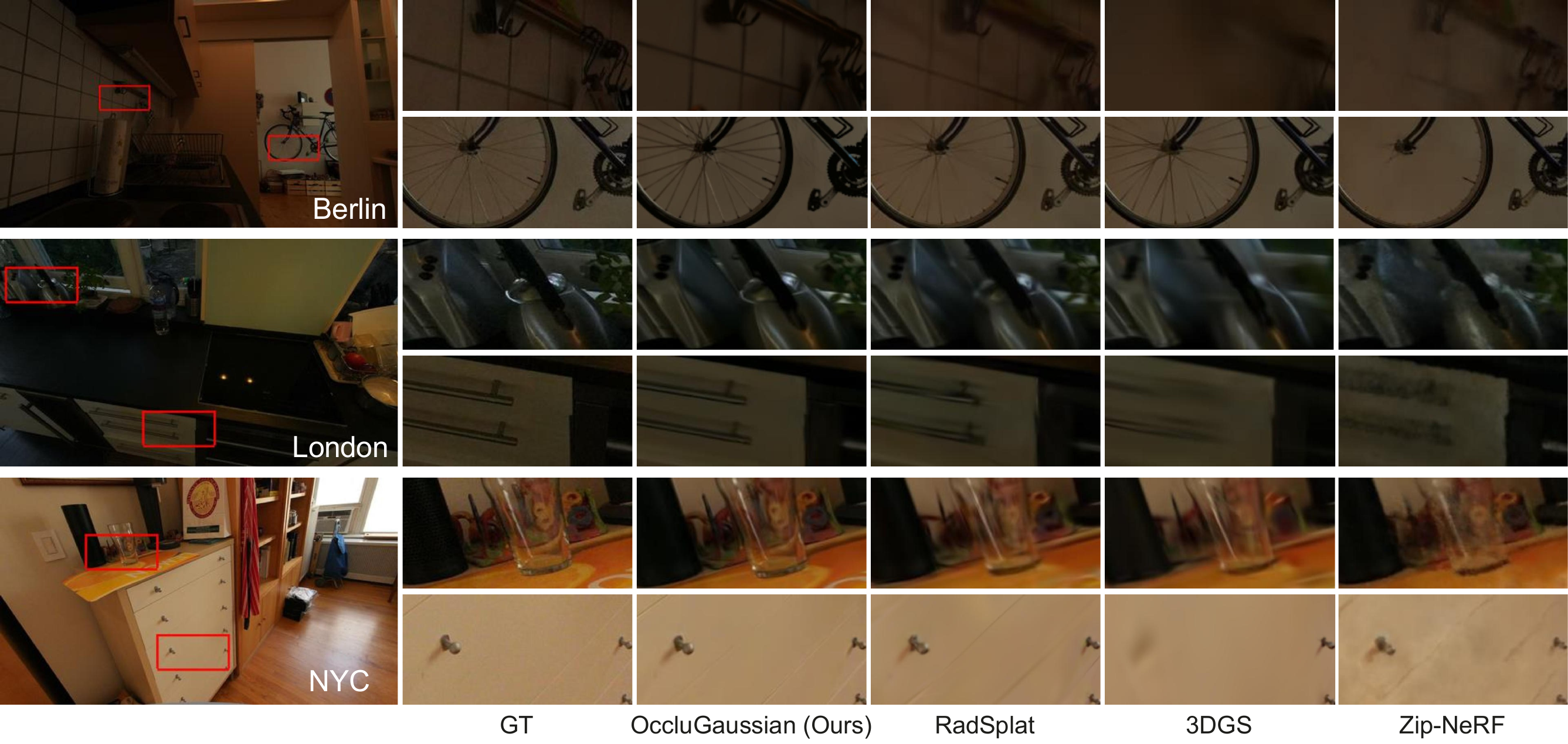}
	\caption{Qualitative comparison with SOTA methods on three large scenes in the Zip-NeRF dataset.}
	\label{fig:visual_comparison_zipnerf}
        \vspace{-3ex}
\end{figure*}
\begin{table*}[t]
  \centering  
  \resizebox{\linewidth}{!}{
    \begin{small}
    \begin{tabular}{l|cccc|cccc|cccc}
    \toprule
     Scene & \multicolumn{4}{c|}{\textsc{Gallery}} & \multicolumn{4}{c|}{\textsc{Canteen}} & \multicolumn{4}{c}{\textsc{ClassBuilding}} \\
    \midrule
    Metrics & PSNR & SSIM & LPIPS & FPS & PSNR & SSIM & LPIPS  & FPS & PSNR & SSIM & LPIPS & FPS \\
    \midrule
    VastGaussian*~\cite{lin2024vastgaussian}  & \underline{25.09} & \textbf{0.903} & \underline{0.095} & 215.22 & \underline{24.60} & \underline{0.890} & \underline{0.105} & 211.02 & \underline{24.05} & \underline{0.884} & \underline{0.111} & 269.97\\
    CityGaussian~\cite{liu2024citygaussian} & 21.98 & 0.808 & 0.294 & 119.86 & 20.41 & 0.794 & 0.275 & 54.02 & 20.48 & 0.840 & 0.244 & 65.57 \\
    Hierarchical-GS~\cite{kerbl2024hierarchical}  & 22.23 & 0.800 & 0.182 & 216.00 & 22.71 & 0.825 & 0.178 & 199.33 & 23.87 & 0.881 & 0.128 & 198.58 \\
    3DGS~\cite{kerbl20233d} & 21.36 & 0.843 & 0.213 & \textbf{344.92} & 21.86 & 0.847 & 0.183 & \textbf{525.93} & 19.41 & 0.871 & 0.186 & \textbf{395.13} \\
    \midrule
    OccluGaussian & \textbf{25.81} & \textbf{0.903} & \textbf{0.094} &  \underline{288.94} & \textbf{25.25} & \textbf{0.900} & \textbf{0.100} &  \underline{311.59} & \textbf{25.33} & \textbf{0.921} & \textbf{0.083} & \underline{339.64} \\
    \bottomrule
    \end{tabular}
    \end{small}
    }
      \caption{Quantitative comparison on the OccluScene3D dataset. We report SSIM $\uparrow$, PSNR $\uparrow$, LPIPS $\downarrow$ and FPS $\uparrow$ on the test views. The \textbf{best} and \underline{second best} results are highlighted. * denotes that it is our re-implementation of VastGaussian.}
    \label{table:quantitative_results}
    \vspace{-1ex}
\end{table*}
\begin{table}[t]
  \centering
  \begin{footnotesize}
    \begin{tabular}{l|ccc}
    \toprule
    Metrics & PSNR & SSIM & LPIPS \\
    \midrule
    MERF~\cite{reiser2023merf}  & 23.49 & 0.747 & 0.445 \\
    SMERF~\cite{duckworth2024smerf}  & 27.28 & 0.829 & 0.340 \\
    Zip-NeRF~\cite{barron2023zip} & 27.37 & 0.836 & 0.305 \\
    \midrule
    3DGS~\cite{kerbl20233d}  & 25.50 & 0.809 & 0.369 \\
    RadSplat~\cite{niemeyer2024radsplat} & 26.17 & 0.839 & 0.364 \\
    \midrule
    \textbf{OccluGaussian} & \textbf{28.63} & \textbf{0.880} & \textbf{0.281} \\
    \bottomrule
    \end{tabular}
    \end{footnotesize}
    \vspace{-1ex}
    \caption{Quantitative comparison on the Zip-NeRF dataset.}
    \label{table:zipnerf_quantitative_results}
\end{table}
\subsection{Region-Based Rendering} \label{sec:rendering}
%
Large scenes contain a substantial number of 3D Gaussians, making rendering computationally expensive and slow. 
Considering that many Gaussian primitives in the scene are occluded, invisible to the viewpoint, culling them in advance can significantly reduce the processing load, thereby accelerating rendering speeds without sacrificing visual quality.
Therefore, leveraging our occlusion-aware scene division, we propose a region-based rendering strategy that culls invisible Gaussians for efficient rendering in large scenes, as shown in the right part of Fig.~\ref{fig:overview}.
%
%

\noindent\textbf{Region-based visibility calculation.}
Consider a large scene reconstructed by $N_g$ 3D Gaussians and divided into $k$ regions $\{R_i\}^{k}_{i=1}$. 
For each region $R_j$, we define a visibility mask $M_j = \{ m_i^j \in \{0, 1\} \}^{N_g}_{i=1}$ for all $N_g$ 3D Gaussians, indicating their visibility from any training viewpoint in $R_j$.
To derive $M_j$, we perform 3DGS rasterization for each camera that is clustered into $R_j$, and calculate each 3D Gaussian $i$'s maximal accumulated weight as its contribution to the rendered image. We mark $m_i^j = 1$ if 3D Gaussian $i$'s contribution to any pixel exceeds 0.01.
Note that we generate both the original view and the backside view from the camera position to comprehensively collect visible 3D Gaussians for each region.
After iterating over all cameras in $R_j$, Gaussians with $m_i^j = 0$ are treated as \emph{occluded} and can be culled in advance when rendering.

\noindent\textbf{Region subdivision.}
Viewpoints near region borders often observe too many 3D Gaussians from neighboring regions, leading to redundant computations when rendering the interior of a region.
To address this issue, we further subdivide each region into smaller sub-regions when deriving visibility masks: one interior sub-region, and several border sub-regions.
First, we identify the boundary lines between the region and its neighboring regions.
Then, we shrink the boundary lines inward by $0.1 d_{max}$, where $d_{max}$ is the maximal distance between two cameras within the region, to obtain the corresponding border sub-regions. 
The remaining area in the region, excluding all the border sub-regions, forms the interior sub-region.
Our region subdivision strategy leads to more compact region-based visibility masks, and thus further boosts the rendering speed.

\noindent\textbf{Rendering with region-based culling.} To render at an arbitrary viewpoint in the complete model, we first identify the region 
 $R_j$ where the viewpoint belongs to, then use the corresponding visibility mask $M_j$ to cull the occluded 3D Gaussians, and finally perform 3DGS rasterization for the remaining 3D Gaussians to generate the rendering image.
By knowing the occlusions between regions, our culling strategy retains only the minimal set of necessary 3D Gaussians for rendering within each region. This approach leads to significant rendering speedup without noticeable quality drop, as shown in the supplementary material.

%% file: sec/4_experiments.tex
\section{Experiments}\label{Experiments}
\subsection{Experimental Setup}
{\bf Dataset.}
Most existing large-scene datasets~\cite{lin2022capturing, li2023matrixcity} are primarily captured from aerial perspectives and thus with rare occlusions.
Therefore, we construct a new dataset called OccluScene3D, containing three large-scale scenes in a campus with complex layouts and significant occlusions: \textsc{Gallery}, \textsc{Canteen} and \textsc{ClassBuilding}.
More details of OccluScene3D are in the supplementary material.
%
%
We also assess our method on the Zip-NeRF dataset~\cite{barron2023zip}, including four large-scale scenes such as apartments and houses. To show the generality of our method, we further conduct experiments on occlusion-free scenes in the Mill-19~\cite{turki2022mega} and UrbanScene3D~\cite{lin2022capturing} datasets.

\noindent{\bf Implementation details.} 
 The models are optimized for 90,000 iterations with densification from iteration 1,500 to 45,000 at intervals of 100.
 We adopt the appearance modeling in VastGaussian~\cite{lin2024vastgaussian} to fit appearance variations across captured images.
%
%
For clustering the view graph, we set a fixed initial cluster number $K=10$.
We iteratively refine the number of clusters until the camera count in each cluster falls within $[M_c - \sigma_cM_c, M_c + \sigma_cM_c]$, where $\sigma_c=0.5$ and $M_c$ is the average camera count across all clusters.
%

\noindent{\bf Baselines.}
On the OccluScene3D dataset, we compare with scalable 3DGS-based methods~\cite{liu2024citygaussian, kerbl2024hierarchical, lin2024vastgaussian, kerbl20233d}.
%
Since VastGaussian~\cite{lin2024vastgaussian} is not open-sourced, we re-implement its scene division strategy and keep all other hyperparameters the same as ours during 3DGS optimization.
For the Zip-NeRF dataset, we compare with methods that have evaluated on it~\cite{reiser2023merf,duckworth2024smerf,barron2023zip,niemeyer2024radsplat}.
Following the evaluation protocol in RadSplat~\cite{niemeyer2024radsplat}, we optimize two models: one for visualization, and another for quantitative analysis, excluding appearance modeling.
We do not use the monocular depth prior in Hierarchical-GS~\cite{kerbl2024hierarchical} for fairness.
%
\begin{table}[t]
  \centering
  \resizebox{\linewidth}{!}{
    \begin{tabular}{l|ccc|ccc}
    \toprule
     Scene & \multicolumn{3}{c|}{\textsc{Mill-19}} & \multicolumn{3}{c}{\textsc{UrbanScene3D}} \\
    \midrule
    Metrics & PSNR & SSIM & LPIPS & PSNR & SSIM & LPIPS \\
    \midrule
    Mega-NeRF~\cite{turki2022mega}  & 23.09 & 0.572 & 0.393 & 24.35 & 0.659 & 0.369 \\
    Switch-NeRF~\cite{zhenxing2022switch}  & 23.50 & 0.590 & 0.355 & 24.84 & 0.678 & 0.333 \\
    Grid-NeRF~\cite{xu2023grid} & 24.37 & 0.807 & 0.142 & 24.94 & 0.787 & 0.158 \\
    \midrule
    Modified 3DGS~\cite{lin2024vastgaussian} & 24.90 & 0.785 & 0.163 & 24.18 & 0.793 & 0.199 \\
    CityGaussian~\cite{liu2024citygaussian} & 24.29 & 0.771 & 0.166 & 23.87 & 0.821 & 0.161 \\
    Hierarchical-GS~\cite{kerbl2024hierarchical} & 22.95 & 0.739 & 0.291 & - & - & -  \\
    DOGS~\cite{chen2025dogs} & 24.26 & 0.762 & 0.231 & 23.40 & 0.742 & 0.280 \\
    VastGaussian~\cite{lin2024vastgaussian} & {25.21} & {0.814} & {0.131} & {25.69} & {0.851} & {0.132} \\
    \midrule
    OccluGaussian & \textbf{25.97} & \textbf{0.854} & \textbf{0.103} & \textbf{25.75} & \textbf{0.858} & \textbf{0.125}  \\
    \bottomrule
    \end{tabular}
    }
  \caption{Quantitative comparison on aerial capture datasets, Mill-19 and UrbanScene3D. 
  We fail to test Hierarchical-GS~\cite{kerbl2024hierarchical} on the \textsc{UrbanScene3D} dataset due to an out-of-memory issue.
  }
    \vspace{-3ex}
    \label{table:urban3d_quantitative_results}
\end{table}

\noindent{\bf Metrics.} We evaluate the rendering quality using three commonly-used quantitative metrics: SSIM, PSNR and LPIPS.
%
We also report the average rendering speed at 1080p resolution in frames per second (FPS).

\subsection{Result Analysis}
\begin{figure}[t]
	\centering
	\includegraphics[width=\linewidth]{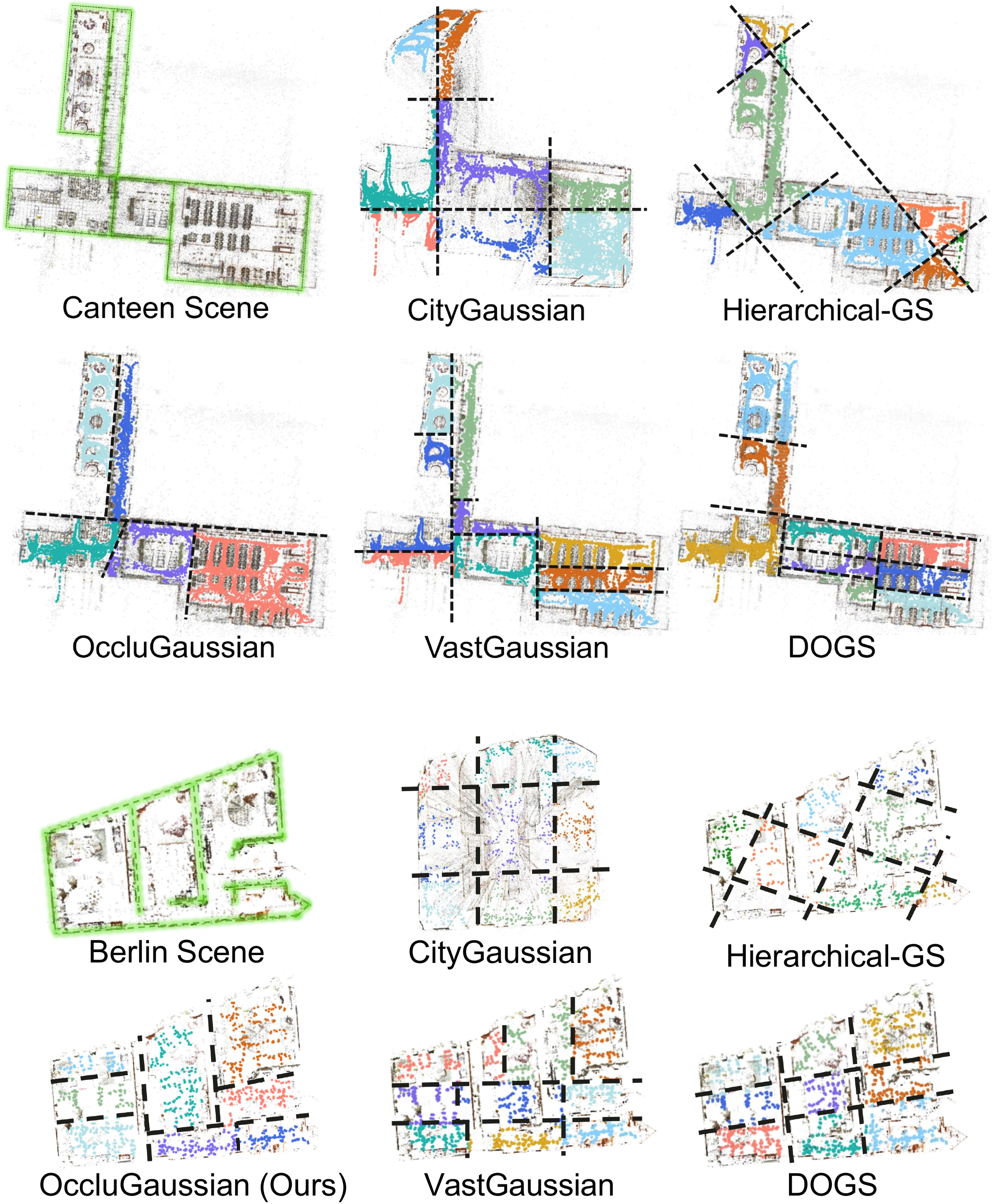}
	\caption{Different division results on the \textsc{Canteen} scene (upper) and \textsc{Berlin} scene (lower). Green lines denote the physical walls, while black lines denote the boundaries of the divided regions. Notably, CityGaussian’s division is projected onto a contracted space.}
    \label{fig:division_results}
    \vspace{-2.5ex}
\end{figure}

\textbf{Reconstruction quality and rendering speeds.} We present quantitative results for occluded scene reconstruction in \cref{table:quantitative_results} and \cref{table:zipnerf_quantitative_results}. We are unable to evaluate DOGS~\cite{chen2025dogs} on the OccluScene3D and Zip-NeRF datasets due to insufficient GPU resources required from its distributed optimization strategy. OccluGaussian almost significantly outperforms the compared methods in all metrics, especially in PSNR. 
Regarding FPS, our method is faster than all methods except the original 3DGS~\cite{kerbl20233d}, due to its less detailed reconstruction and fewer Gaussian primitives.
Visual comparisons in \cref{fig:visual_comparison} and \cref{fig:visual_comparison_zipnerf} illustrate OccluGaussian's superior detail in rendering. 
Additionally, \cref{fig:division_results} shows OccluGaussian's scene division aligns better with the scene layout. More visualizations are provided in the supplementary material.

In \cref{table:urban3d_quantitative_results}, we validate our method on occlusion-free scenes using the Mill-19~\cite{turki2022mega} and UrbanScene3D~\cite{lin2022capturing} datasets. Our method maintains a clear advantage, highlighting the generality of our approach.
These results show the effectiveness of our scene division strategy, which enhances the correlation among training cameras within each region and improves reconstruction quality.

\subsection{Comparison of Camera Clustering Methods }
\begin{figure}[t]
	\centering
	\includegraphics[width=\linewidth]{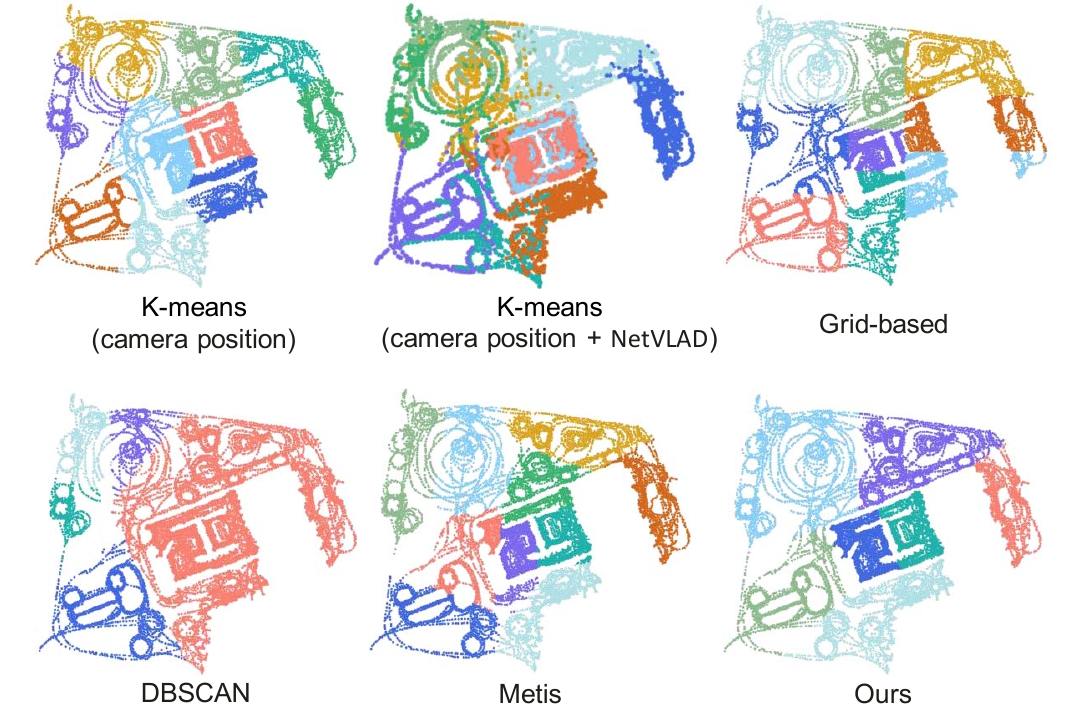}
	\caption{Scene division results by different methods are presented with points for cameras, color-coded by their respective regions.}
    \label{fig:clutering_methods_comparison}
    \vspace{-1ex}
\end{figure}

We compare various camera clustering methods on the \textsc{Gallery} scene in the OccluScene3D dataset.
As shown in \cref{fig:clutering_methods_comparison}, grid-based methods partition scenes into grids without considering clustering features and occlusions. 
The K-means algorithm, which relies solely on camera position features, is also oblivious to occlusions and performs worse when incorporating image features extracted by NetVLAD~\cite{arandjelovic2016netvlad}, failing to extract region boundaries.
DBSCAN~\cite{ester1996density} does not require an initial cluster number, but produces unevenly sized clusters and also lacks occlusion awareness.
Metis~\cite{karypis1997metis} provides slightly better results by conducting graph clustering, but severe occlusions persist in the partitioned regions.
We further perform 3DGS optimization based on these clustering results, as shown in~\cref{table:quantitative_different_cluster_method}, which demonstrates that our clustering method outperforms others in all metrics.


\begin{table}[t]
        \scriptsize
	\centering
 \resizebox{\linewidth}{!}{
	\begin{tabular}{lccc}
		\toprule
		Method & PSNR$\uparrow$ & SSIM$\uparrow$ & LPIPS$\downarrow$ \\
		\midrule
  		Grid-based & 24.58 & 0.892 & 0.104 \\
            \midrule
		  K-means (CamP) & 24.94 & 0.903 & 0.094 \\
            K-means (CamPN) & - & - & - \\
		  Metis~\cite{karypis1997metis} & 25.02 & 0.903 & 0.097  \\
            DBSCAN~\cite{ester1996density} & 25.00 & 0.898 & 0.111   \\
            \midrule
            Ours & \textbf{25.46} & \textbf{0.908} & \textbf{0.094}  \\ 
		\bottomrule
	\end{tabular}
}
\vspace{-2ex}
    \caption{Comparison of clustering methods. `CamP' denotes camera position, `CamPN' denotes camera position + NetVLAD. K-means (CamPN) fails due to its poor clustering results that prevent effective distinctions by the classification model.}
    \label{table:quantitative_different_cluster_method}
\end{table}
\begin{table}[t]
        \scriptsize
	\centering
 \resizebox{\linewidth}{!}{
	\begin{tabular}{lccccc}
		\toprule
		& PSNR & SSIM & LPIPS & FPS \\
		\midrule
        w/o RBR \& RSD & 25.81 & 0.903 & 0.099 & 189.52  \\
		w/o RSD & 25.81 & 0.903 & 0.099 & 271.79 \\
		Full  & \textbf{25.81} & \textbf{0.903} & \textbf{0.099} & \textbf{288.94} \\ 
		\bottomrule
	\end{tabular}
 }
  \vspace{-2ex}
 	\caption{Ablation study of the region-based rendering on the \textsc{gallery} scene. `RBR' denotes vanilla region-based rendering; `RSD' denotes region subdivision in our region-based rendering.}
    \label{table:ablation_culling}
  \vspace{-5ex}
\end{table}

\subsection{Ablation Study}
\textbf{Region-based rendering.} We ablate our region-based rendering on the \textsc{gallery} scene, as shown in \cref{table:ablation_culling}.
Our region-based culling strategy significantly enhances rendering speeds without noticeable loss in visual quality. Additionally, the proposed region subdivision technique further accelerates rendering. 
It is important to note that our culling strategy automatically removes Gaussian primitives with minimal contribution, reducing the overall number of Gaussian primitives in the final reconstructed model.

\noindent\textbf{Division strategy.} To demonstrate the effectiveness of our occlusion-aware scene division, we evaluate it alongside VastGaussian, CityGaussian, Hierarchical-GS, and DOGS by replacing our scene division with theirs while keeping the same hyperparameters as ours for 3DGS optimization. We test on the \textsc{Canteen} scene of OccluScene3D dataset and the \textsc{Berlin} scene of Zip-NeRF. Results in \cref{tab:ablation_clustering_type_strategy} shows that our occlusion-aware scene division still demonstrates a significant advantage. Additional visualizations can be found in the supplementary material.

\noindent\textbf{Initial clustering numbers.} To show the robustness of our method, we vary the manually-set initial clustering number $K$ on the \textsc{NYC} scene of Zip-NeRF dataset and find similar performance trends, as shown in \cref{tab:ablation_K}. However, for huge scenes with dozens or hundreds of separated regions, setting an appropriate initial clustering number $K$ remains crucial, which we leave for future exploration.

\begin{table}[t]
\begin{center}
\resizebox{1\linewidth}{!}{
\begin{tabular}{ccccccc}
\toprule
 & \multicolumn{3}{c}{Canteen} & \multicolumn{3}{c}{Berlin} \\
 \cmidrule(lr){2-4} \cmidrule(lr){5-7}
 & PSNR & LPIPS & \#Blocks  & PSNR & LPIPS & \#Blocks \\
\midrule
VastGaussian~\cite{lin2024vastgaussian} & 24.60 & 0.105 & 9 & 29.48 & 0.085 & 9 \\
CityGaussian~\cite{liu2024citygaussian} & 24.16 & 0.114 & 9 & 28.68 & 0.091 & 9 \\
Hierarchical-GS~\cite{kerbl2024hierarchical} & 23.68 & 0.131 & 9 & 27.26 & 0.105 & 9 \\
DOGS~\cite{chen2025dogs} & 24.66 & 0.108 & 9 & 28.15 & 0.095 & 9 \\
\midrule
OccluGaussian & \textbf{25.25} & \textbf{0.100} & 5 & \textbf{30.37} & \textbf{0.076} & 8 \\
\bottomrule
\end{tabular}
}
\end{center}
\vspace{-3ex}
\caption{Different division strategies used in OccluGaussian.}
\label{tab:ablation_clustering_type_strategy}
\end{table}

\begin{table}[t]
\begin{center}
\resizebox{0.8\linewidth}{!}{
\begin{tabular}{ccccc}
\toprule
Initial $K$ & Final $K$ & PSNR & SSIM & LPIPS \\
\midrule
7 & 6 & 30.90 & 0.898 & 0.126  \\
10 & 7 & 31.33 & 0.902 & 0.121  \\
15 & 8 & 31.35 & 0.901 & 0.121  \\
\bottomrule
\end{tabular}
}
\end{center}
\vspace{-3ex}
\caption{Different initial clustering numbers $K$ on the NYC scene of Zip-NeRF dataset.}
\label{tab:ablation_K}
\vspace{-3ex}
\end{table}

%% file: sec/5_conclusion.tex
\section{Conclusion}
This paper presents OccluGaussian, a novel approach for high-quality reconstruction and real-time rendering of large-scale scenes.
We propose an occlusion-aware scene division strategy that enhances reconstruction quality by optimizing the contribution of training cameras within each region.
Furthermore, we propose a region-based rendering strategy that discards occluded 3D Gaussians, significantly accelerating rendering for large-scale scenes.

\noindent\textbf{Limitation.} While OccluGaussian has made significant progress, limitations remain. For instance, our camera clustering starts with a fixed initial number, which works well in our experiments but may be insufficient for extremely large scenes. Future work will focus on dynamically determine the initial clustering number based on the number of training cameras to enhance robustness and scalability.

%% file: sec/X_suppl.tex
\clearpage
\setcounter{page}{1}
\maketitlesupplementary

\section{Training details}
All experiments were conducted on RTX 4090 GPUs, except for Hierarchical-GS on V100s due to higher memory demand. 
The reported time includes only 3DGS optimization, excluding preprocessing steps. Our division strategy takes an average of 9.28 minutes on the OccluScene3D dataset and can be greatly accelerated by porting it from CPU to GPU in future work. For fair comparison, we used consistent settings for all methods based on their public codes (except DOGS, which requires distributed training beyond our GPU capacitance) on the MILL-19 and URBANSCENE3D datasets: dividing each scene into 9 blocks, downsampling all images by 4, and optimizing each block for 60k steps with 200-step densification interval

\section{More Details of the OccluScene3D Dataset}
There are three scenes in the OccluScene3D dataset:
\textsc{Gallery}, \textsc{Canteen} and \textsc{ClassBuilding}.
All the videos are recorded by a mobile phone with the wide-angle mode and landscape orientation at a frame rate of 60 Hz.
We use COLMAP~\cite{schonberger2016structure} to estimate the camera intrinsic and extrinsic parameters.
More details are shown in \cref{table:dataset_details}. 

\begin{figure}[t]
	\centering
	\includegraphics[width=\linewidth]{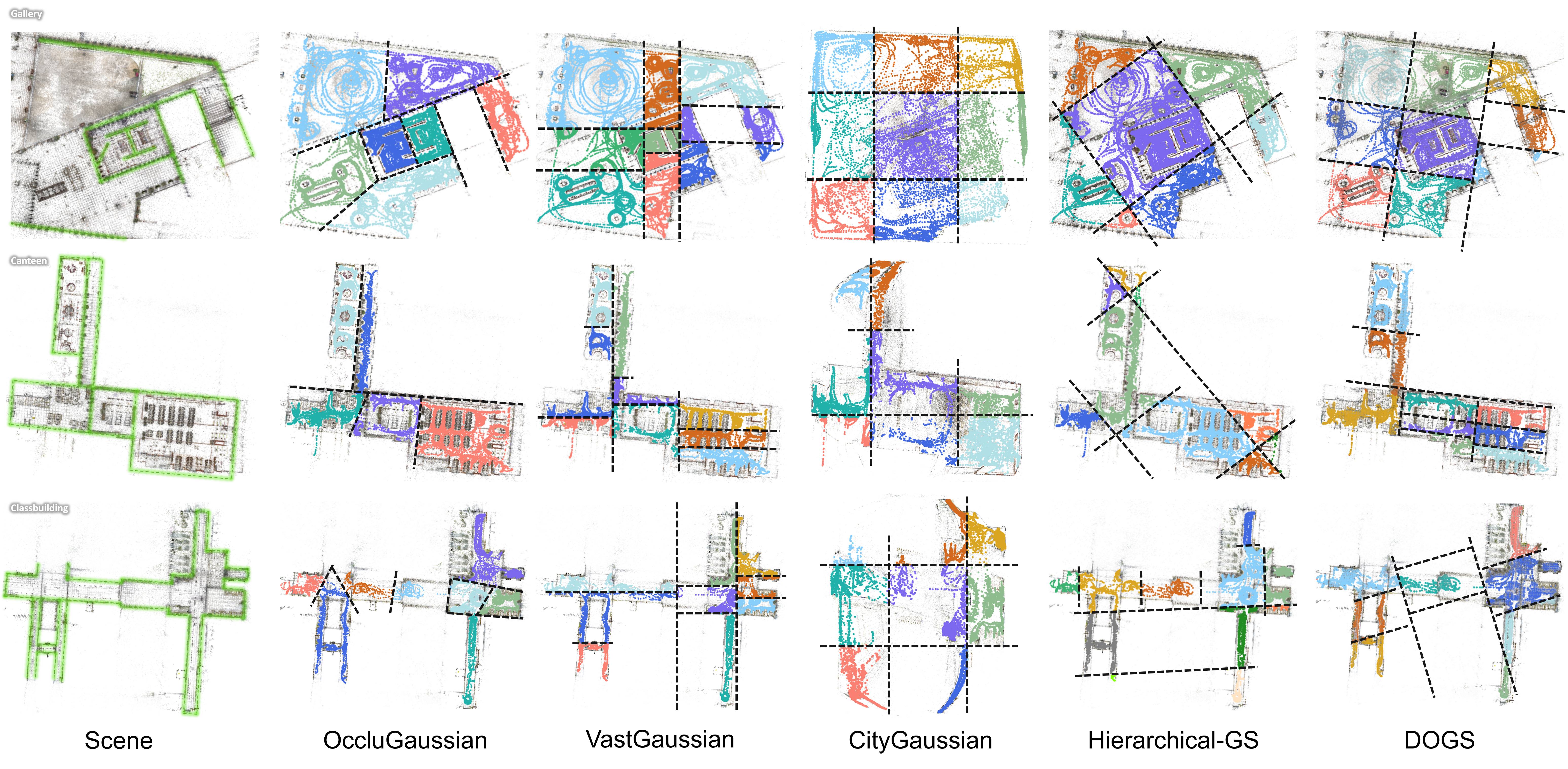}
	\caption{Comparative analysis of scene division on \textsc{Gallery}, \textsc{Canteen} and \textsc{ClassBuilding} in the OcclusionScene3D dataset. Notably, CityGaussian's division is projected onto a contracted space.}
	\label{fig:division_results_OccluScene3D}
\end{figure}

\begin{figure}[t]
	\centering
	\includegraphics[width=\linewidth]{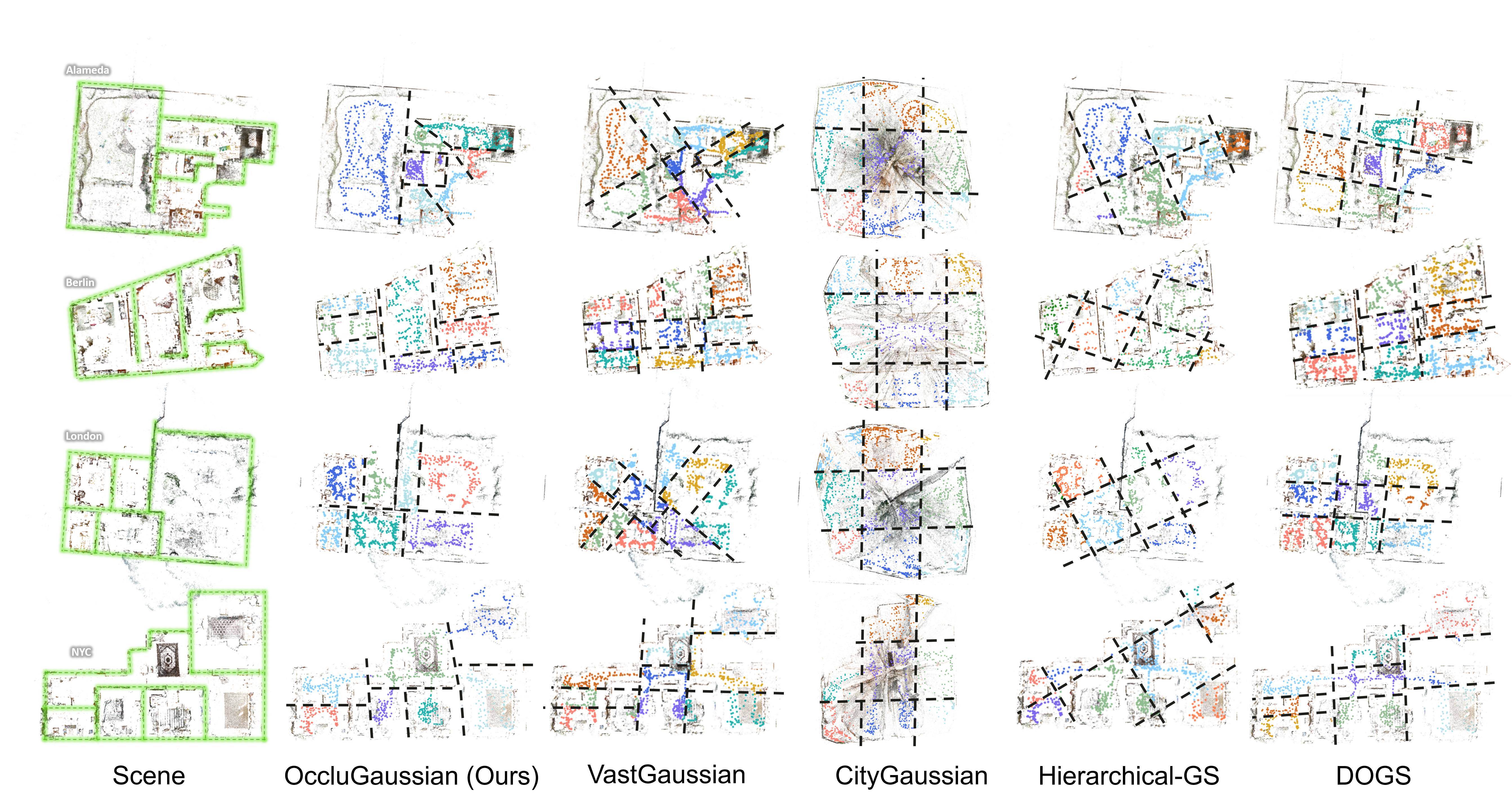}
	\caption{Comparative analysis of scene division on \textsc{Alameda}, \textsc{Berlin}, \textsc{London} and \textsc{NYC} in the Zip-NeRF dataset. Notably, CityGaussian's division is projected onto a contracted space.}
	\label{fig:zipnerf_division}
\end{figure}

\begin{table}[t]
  \centering
  \scriptsize
  \begin{tabular}{lrrrr}
    \toprule
    Scene & Area ($m^2$) & \#Video & Duration ($min$) & \#Image \\
    \midrule
    \textsc{Gallery} & 2500 & 71 & 174 & 9881   \\
    \textsc{Canteen} & 1500 & 46 & 111 & 9034   \\
    \textsc{ClassBuilding} & 1000 & 18 & 76 & 9118  \\
    \bottomrule
  \end{tabular}
    \caption{Statistics of the OccluScene3D dataset with three real scenes. Area: covered area; \#Video: total number of videos; Duration: total duration of all the videos; \#Image: total number of sampled images for reconstruction.}
  \label{table:dataset_details}
\end{table}

\begin{table}[t]
\begin{center}
\resizebox{1\linewidth}{!}{
\begin{tabular}{ccccccc}
\toprule
 & PSNR & Train (m) & Mem (GB) & \#GS (M) & FPS \\
\midrule
VastGaussian & 24.58 & \underline{49.97} & 15.0 & 5.5 & 232 \\
CityGaussian & 20.96 & 106.35 & \textbf{8.2} & 6.4 & 80 \\
Hierarchical-GS & 22.94 & 190.36 & 30.5 & 40.0 & 205 \\
3DGS & 20.88 & 86.8 & 23.9 & \textbf{0.9} & \textbf{422} \\
\midrule
OccluGaussian & \textbf{25.46} & \textbf{48.26} & \underline{12.2} & \underline{3.4} & \underline{313} \\
\bottomrule
\end{tabular}
}
\end{center}
\vspace{-15pt}
\caption{More quantitative results on OccluScene3D dataset.}\label{tab:training_time}
\end{table}

\begin{table}[ht]
  \centering
  \resizebox{\linewidth}{!}{
    \begin{tabular}{l|ccc|ccc}
    \toprule
     Scene & \multicolumn{3}{c|}{\textsc{Garden}} & \multicolumn{3}{c}{\textsc{Kitchen}} \\
    \midrule
    Metrics & PSNR & SSIM & LPIPS & PSNR & SSIM & LPIPS \\
    \midrule
    Vanilla 3DGS & \textbf{27.41} & 0.868 & 0.103 & 30.31 & 0.922 & 0.12 \\
    Ours (w/o division) & 27.32 & \textbf{0.879} & 0.068 & \textbf{30.78} & 0.922 & 0.065 \\
    Ours (two-block division) & 27.37 & \textbf{0.879} & \textbf{0.067} & 30.73 & \textbf{0.923} & \textbf{0.064} \\
    \bottomrule
    \end{tabular}
    }
  \vspace{-1ex}
  \caption{Quantitative comparison on the Mip-NeRF 360 dataset.
  }
    \label{table:Mip-NeRF_360_results}
    \vspace{-4ex}
\end{table}

\begin{figure*}[t]
	\centering
	\includegraphics[width=\linewidth]{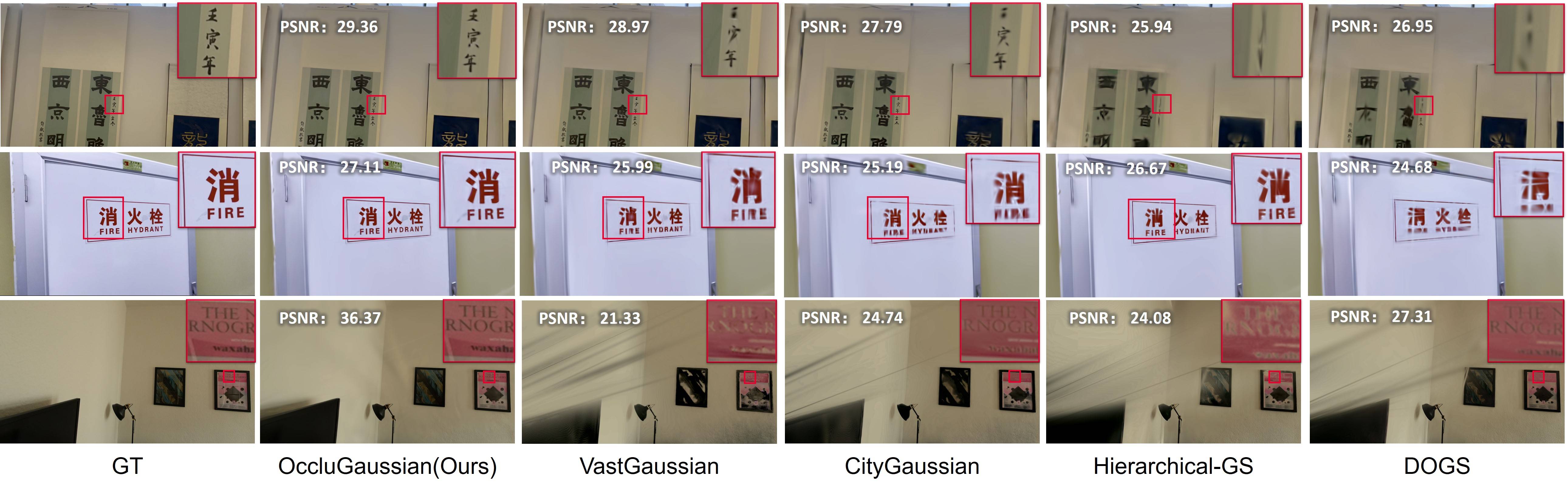}
	\caption{Quantitative evaluation of different division strategies applied under the same 3DGS optimization hyperparameters. Lines from top to bottom are the \textsc{Gallery}, and \textsc{Classbuilding} scenes from the OccluScene3D dataset, as well as the \textsc{Berlin} scene from the Zip-NeRF dataset.}
	\label{fig:different_division_method_on_occlugaussian}
\end{figure*}

\begin{table*}[t]
  \centering
  \resizebox{\textwidth}{!} {
    \begin{tabular}{l|ccc|ccc|ccc|ccc|ccc}
    \toprule
     \multirow{2}{*}{Scene} & \multicolumn{6}{c|}{\textsc{Mill-19}} & \multicolumn{9}{c}{\textsc{UrbanScene3D}} \\
      \cmidrule{2-16}
      & \multicolumn{3}{c|}{\textsc{Building}} & \multicolumn{3}{c|}{\textsc{Rubble}} & \multicolumn{3}{c|}{\textsc{Campus}} & \multicolumn{3}{c|}{\textsc{Residence}} & \multicolumn{3}{c}{\textsc{Sci-Art}}\\
    \midrule
    Metrics & PSNR & SSIM  & LPIPS & PSNR & SSIM  & LPIPS & PSNR & SSIM  & LPIPS & PSNR & SSIM  & LPIPS & PSNR & SSIM  & LPIPS \\
    \midrule
    VastGaussian~\cite{lin2024vastgaussian}  & \underline{23.50} & \underline{0.804} & \underline{0.130} & \underline{26.92} & \underline{0.823} & \underline{0.132} & \underline{26.00} & \underline{0.816} & \underline{0.151} & \textbf{24.25} & \textbf{0.852} & \underline{0.124} & \textbf{26.81} & \underline{0.885} & \underline{0.121} \\
    CityGaussian~\cite{liu2024citygaussian}  & 22.59 & 0.757 & 0.174 & 25.98 & 0.784 & 0.158 & - & - & - & 23.25 & 0.806 & 0.156 & 24.49 & 0.836 & 0.167 \\
    Hierarchical-GS~\cite{kerbl2024hierarchical}  & 21.25 & 0.723 & 0.297 & 24.64 & 0.755 & 0.284 & - & - & - & - & - & - & - & - & -\\
    3DGS~\cite{kerbl20233d} & 23.01 & 0.769 & 0.164 & 26.78 & 0.800 & 0.161 & 23.89 & 0.712 & 0.289 & 23.40 & 0.825 & 0.142 & 25.24 & 0.843 & 0.166 \\
    \midrule
    OccluGaussian~\cite{liu2024citygaussian} & \textbf{24.77} & \textbf{0.853} & \textbf{0.100} & \textbf{27.16} & \textbf{0.854} & \textbf{0.105} & \textbf{26.60} & \textbf{0.836} & \textbf{0.139} & \underline{24.24} & \underline{0.846} & \textbf{0.122} & \underline{26.42} & \textbf{0.890} & \textbf{0.113}\\
    \bottomrule
    \end{tabular}
    }
      \caption{Quantitative evaluation of our method compared to previous work on the Mill-19~\cite{turki2022mega} and UrbanScene3D~\cite{lin2022capturing} datasets. The
\textbf{best} and \underline{second best} results are highlighted. Due to an out-of-memory issue, we were unable to test Hierarchical-GS [22] on the \textsc{URBANSCENE3D} dataset and CityGaussian on the CAMPUS scene.}
    \label{table:quantitative_results_occlusion_free_detail}
\end{table*}

\begin{table*}[t]
  \centering
  \resizebox{\textwidth}{!} {
    \begin{tabular}{l|ccc|ccc|ccc|ccc}
    \toprule
      & \multicolumn{3}{c|}{\textsc{Berlin}} & \multicolumn{3}{c|}{\textsc{London}} & \multicolumn{3}{c|}{\textsc{NYC}} & \multicolumn{3}{c}{\textsc{Alameda}} \\
    \midrule
    Metrics & PSNR & SSIM  & LPIPS & PSNR & SSIM  & LPIPS & PSNR & SSIM  & LPIPS & PSNR & SSIM & LPIPS \\
    \midrule
    3DGS~\cite{kerbl20233d} & 28.52 & 0.887 & 0.325 & 27.05 & 0.829 & 0.342 & 28.21 & 0.844 & 0.321 & 25.35 & 0.758 & 0.37 \\
    SMERF~\cite{duckworth2024smerf} & 28.52 & 0.887 & 0.325 & 27.05 & 0.829 & 0.342 & 28.21 & 0.844 & 0.321 & 25.35 & 0.758 & 0.37 \\
    Zip-NeRF~\cite{barron2023zip}  & \underline{28.59} & \underline{0.891} & \underline{0.297} & \underline{27.06} & \underline{0.835} & \underline{0.304} & \underline{28.42} & \underline{0.850} & 0.281 & \textbf{25.41} & \underline{0.767} & \underline{0.338} \\
    \midrule
    OccluGaussian & \textbf{30.37} & \textbf{0.937} & \textbf{0.076} & \textbf{28.06} & \textbf{0.868} & \textbf{0.141} & \textbf{31.33} & \textbf{0.902} & \textbf{0.121} & \underline{24.75} & \textbf{0.814} & \textbf{0.201} \\
    \bottomrule
    \end{tabular}
    }
  \caption{Quantitative evaluation of our method compared to previous work on the Zip-NeRF~\cite{barron2023zip} datasets. }
    \label{table:quantitative_results_zipne_detail}
\end{table*}

\begin{table}[t]
	\centering
        \resizebox{\linewidth}{!}{
	\begin{tabular}{lcccccccccc}
		\toprule
		& PSNR & Base cam. & Extended cam. & Ratio \\
		\midrule
		VastGaussian~\cite{lin2024vastgaussian} & 24.58 & 9119 & 8921 & 97.8\%  \\
		CityGaussian~\cite{liu2024citygaussian} & 22.50 & 9119 & 5371 & 58.90\%   \\
		Hierarchical-GS~\cite{kerbl2024hierarchical}& 23.93 & 9119 & 6808 & 74.66\%      \\
            DOGS~\cite{chen2025dogs}& 25.11 & 9119 & 8331 & 91.36\%      \\
            \midrule
            OccluGaussian& \textbf{25.81} & 9119 & \textbf{4232} & \textbf{46.41}\%   \\
		\bottomrule
	\end{tabular}
        }
        \caption{
		  Extended camera comparison on OccluScene3D.
	}
	\label{table:extended_camera_ratio}
\end{table}

\begin{figure}[t]
	\centering
	\includegraphics[width=\linewidth]{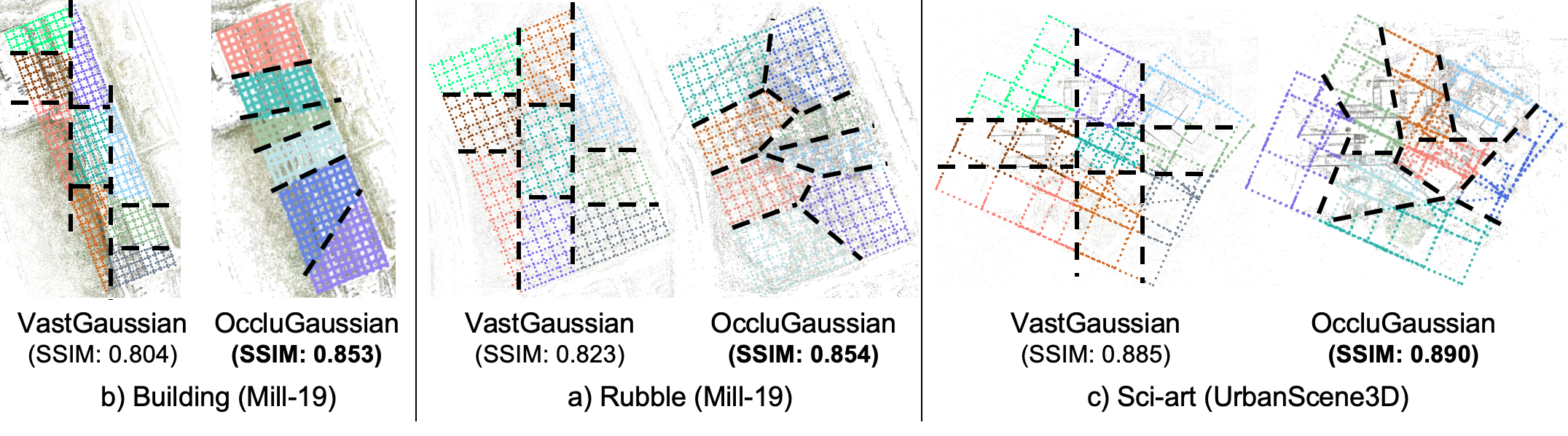}
	\caption{Different division results on aerial datasets.}
        \vspace{-3ex}
	\label{fig:occlusion_free_comparison2}
\end{figure}

\begin{table}[t]
\begin{center}
\resizebox{1\linewidth}{!}{
\begin{tabular}{cccccc}
\toprule
 & \multicolumn{2}{c}{\#Clusters} & \multicolumn{3}{c}{Metrics} \\
 \cmidrule(lr){2-3} \cmidrule(lr){4-6}
 & Initial $K$ & Final $K$ & PSNR & SSIM & LPIPS \\
\midrule
Fixed cluster number & 10 & 10 & 28.092 & 0.870 & 0.141 \\
$M_c\pm0.3M_c$ & 10 & 8 & 28.090 & 0.870 & 0.141 \\
$M_c\pm0.5M_c$ & 10 & 7 & 28.060 & 0.868 & 0.141 \\
$M_c\pm0.7M_c$ & 10 & 6 & 28.043 & 0.862 & 0.142 \\
\bottomrule
\end{tabular}
}
\end{center}
\vspace{-16pt}
\caption{Quantitative results of different camera count ranges.}\label{tab:ablation_clustering_number_strategy}
\end{table}

\section{Additional Experimental Results}
\subsection{Additional Quantitative Analysis}
\noindent{\bf Scene division strategy analysis.} We present a comparison of scene division on the OccluScene3D dataset and the Zip-NeRF dataset among VastGaussian~\cite{lin2024vastgaussian}, CityGaussian~\cite{liu2024citygaussian}, Hierarchical-GS~\cite{kerbl2024hierarchical}, DOGS~\cite{chen2025dogs}, and OccluGaussian in~\cref{fig:division_results_OccluScene3D} and~\cref{fig:zipnerf_division}. 
The results demonstrate that the scene divisions  produced by OccluGaussian better align with scene layouts.
Consequently, OccluGaussian achieves superior reconstruction quality, which is proven by performing 3DGS optimization with the same hyperparameters across different scene division strategies, as illustrated in~\cref{fig:different_division_method_on_occlugaussian}.
For aerial datasets without severe occlusions, our method remains effective. Unlike previous methods using fixed layouts (\eg, 3×3 grid), our division approach is not strictly grid-based. Thanks to our view graph clustering, it can adapt to the distribution of cameras. This enables a more balanced number of cameras across blocks, which is crucial for higher performance, as also demonstrated by CityGaussian and DOGS. Such benefit is especially noticeable in elongated capturing scenarios (\eg, \textsc{Building}, \textsc{Rubble}), but less so in square scenes (\eg, \textsc{Sci-Art}), as illustrated in \cref{fig:occlusion_free_comparison2}. 

We further compare vanilla 3DGS with our method on two scenes from the Mip-NeRF 360 dataset. We evaluate our method with both no-division and two-block division settings.
As these scenes are relatively small and occlusion-free, the divided blocks contain nearly all cameras, showing little difference compared to vanilla 3DGS (see \cref{table:Mip-NeRF_360_results}).
%

\noindent{\bf Detailed quantitative analysis of OccluGaussian.} 
We present the average PSNR, training time, allocated memory, number of gaussians and FPS on OccluScene3D dataset in \cref{tab:training_time}.
We validate the methods on the Mill-19~\cite{turki2022mega}, UrbanScene3D~\cite{lin2022capturing} and Zip-NeRF~\cite{barron2023zip} datasets in \cref{table:quantitative_results_occlusion_free_detail} and \cref{table:quantitative_results_zipne_detail}. It can be observed that OccluGaussian outperforms others in terms of LPIPS among all the datasets. It also overall holds a clear advantage over existing methods in other metrics, highlighting its generality.
These results show the effectiveness of our scene division strategy, which strengthens the correlations among training cameras within each region, and achieves a higher average contribution to the reconstruction results. 

\begin{figure}[t]
	\centering
\includegraphics[width=1\linewidth]{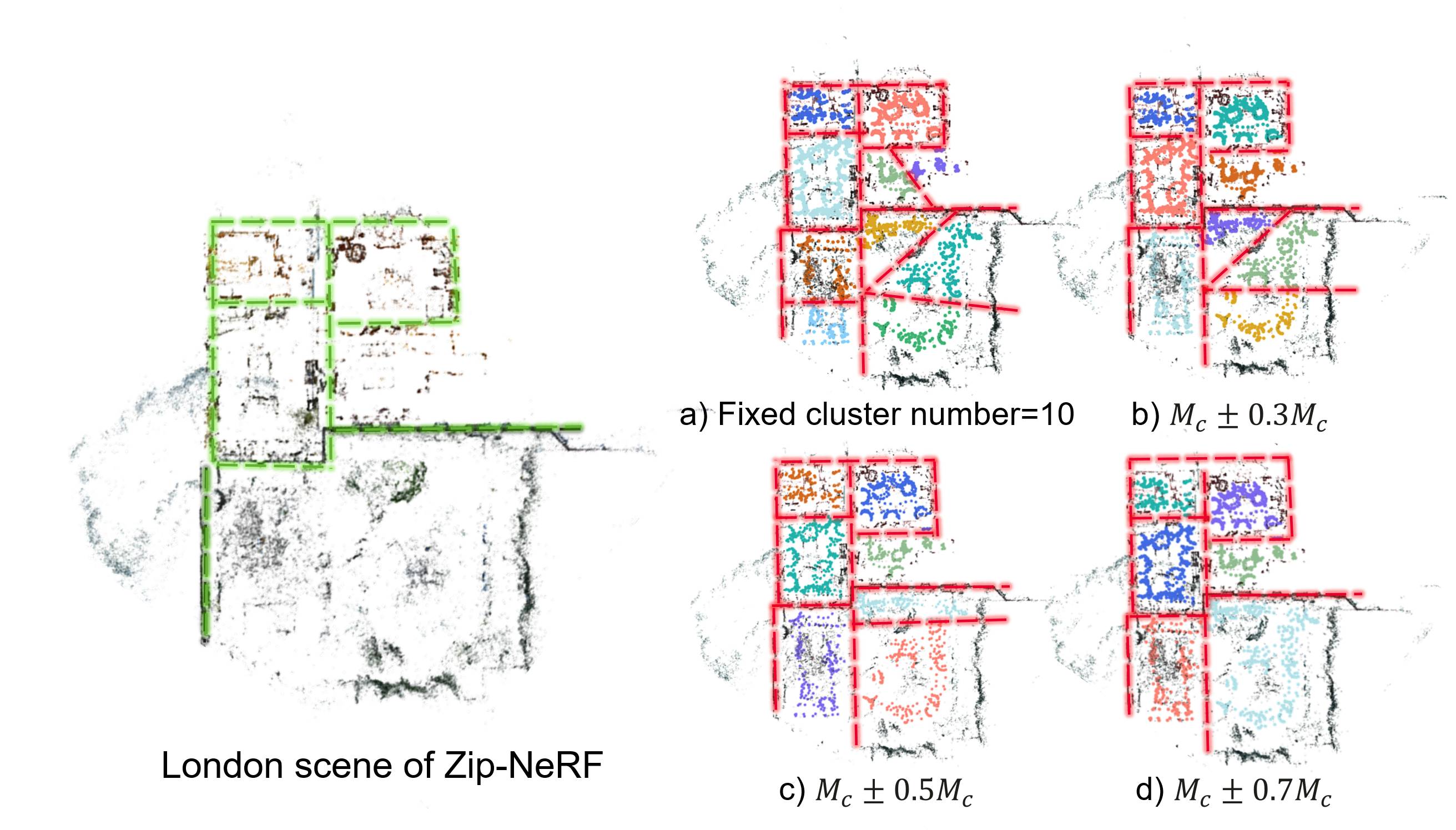}
	\caption{Division results of different camera count ranges for each cluster. The green lines denote the physical walls, and the red lines denote the boundaries of the divided regions.}
 \label{fig:ablation_clustering_number_strategy}
 \vspace{-2ex}
\end{figure}

\begin{figure}[t]
	\centering
	\includegraphics[width=\linewidth]{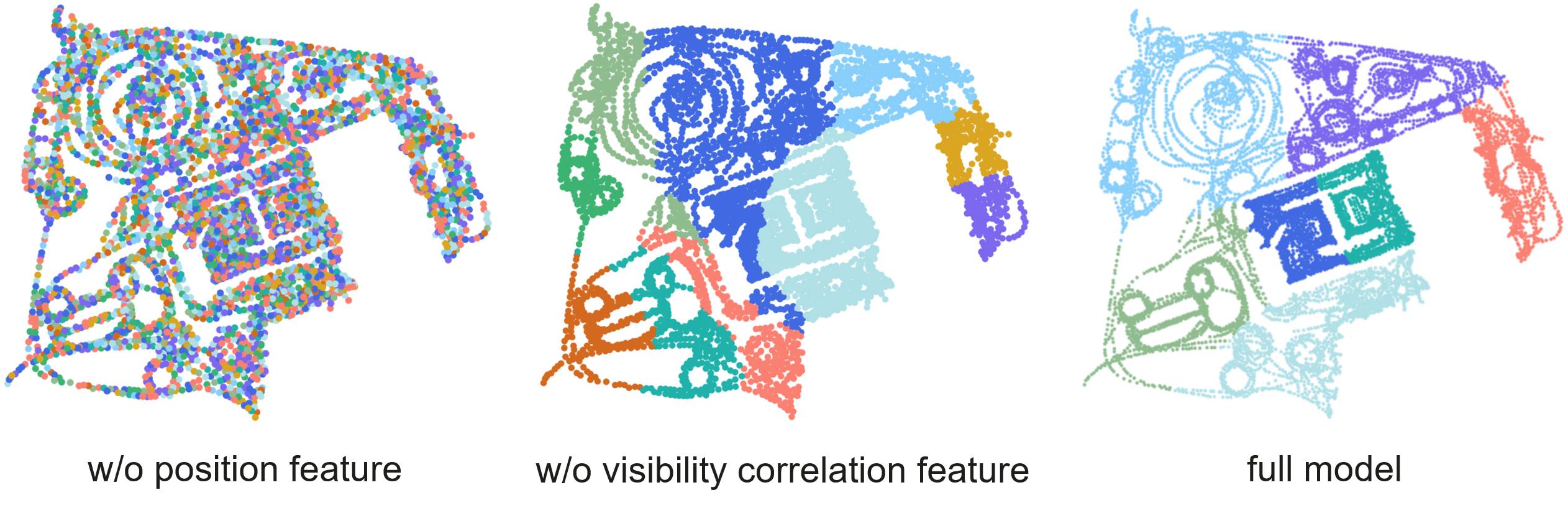}
	\caption{Ablation study of features in our attributed scene graph.}
	\label{fig:clustering_features_ablation}
    \vspace{-2ex}
\end{figure}

\begin{figure}[t]
	\centering
	\includegraphics[width=\linewidth]{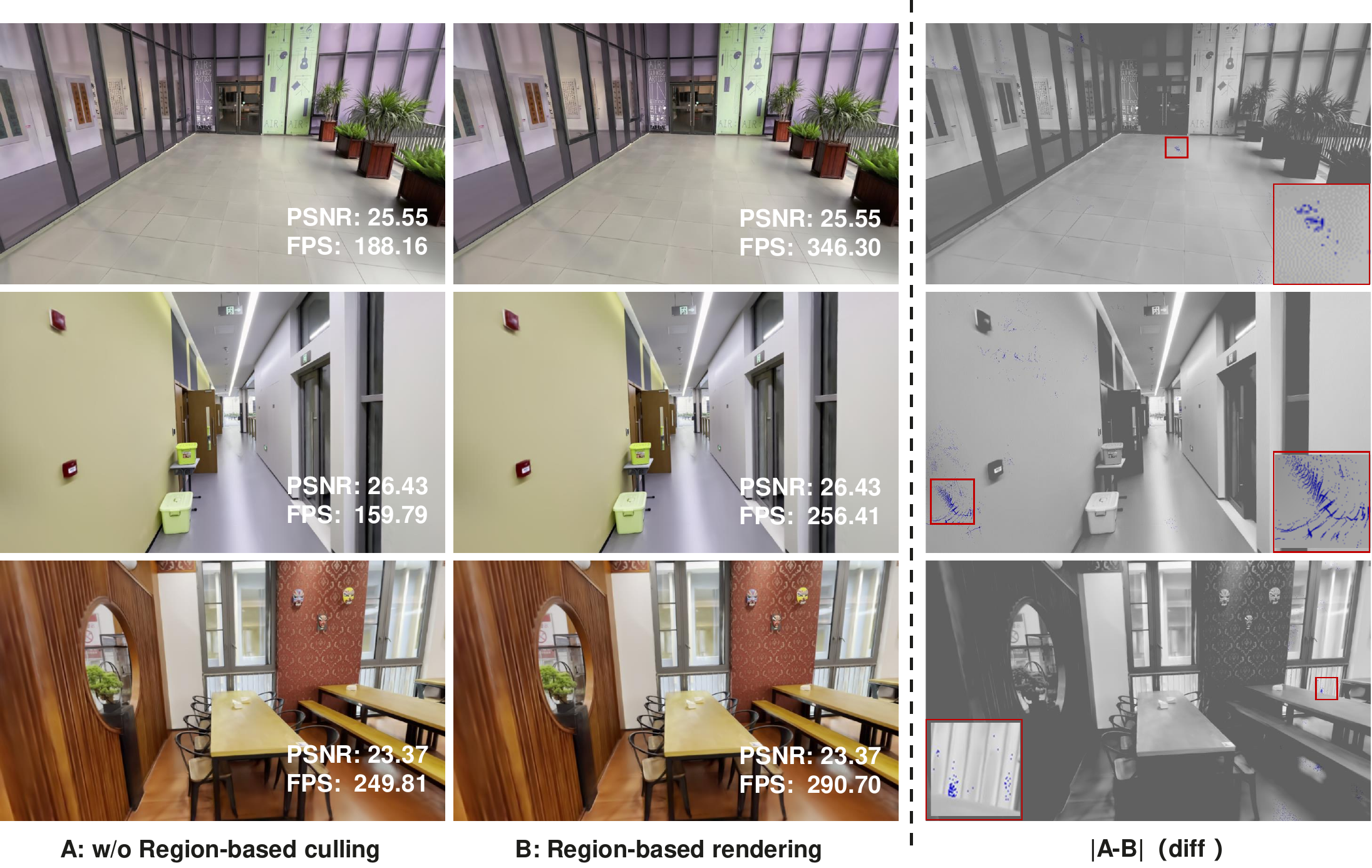}
	\caption{Quantitative evaluation of region-based rendering. The right-most image uses color coding to represent pixel differences: red for $\left|A-B\right| > 1$, blue for $\left|A-B\right| = 1$, and gray for A-B = 0. The difference between A and B is almost less than 1 pixel. Our approach achieves substantial enhancements in rendering speed without a perceptible loss in image quality.}
	\label{fig:ablation_culling_figure}
\end{figure}

\begin{figure}[t]
	\centering
	\includegraphics[width=\linewidth]{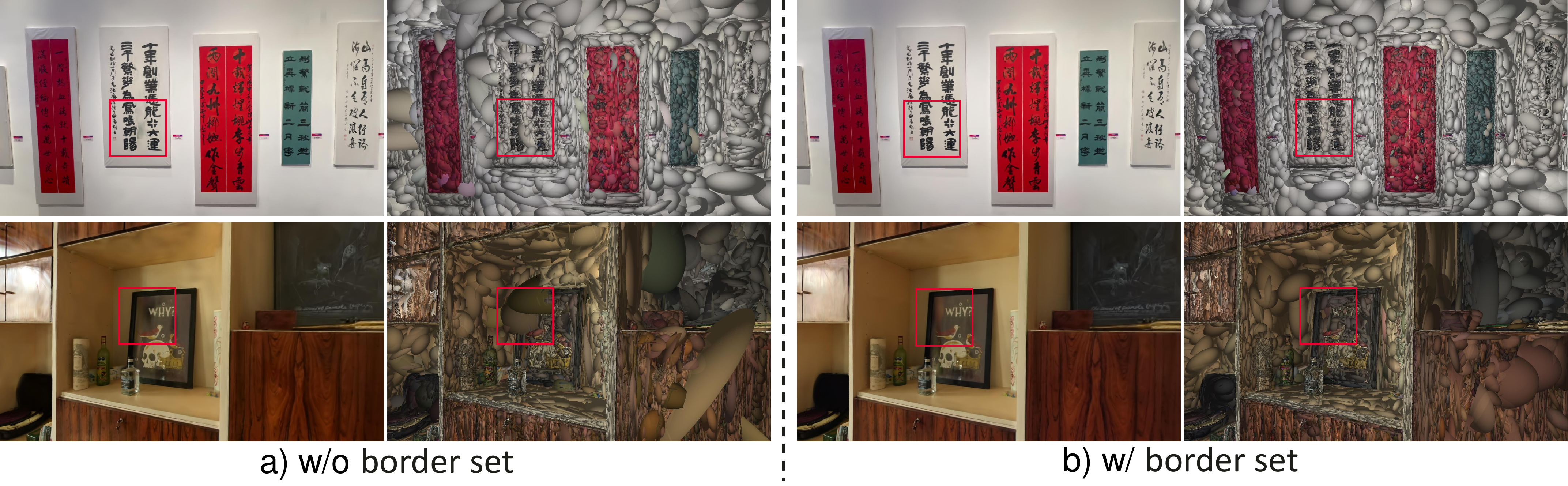}
	\caption{{Ablation study of border set cameras on GALLERY and BERLIN. We also report the performance on the test sets.}}
	\label{fig:ablation study of border set}
\end{figure}

\begin{figure}[t]
	\centering
	\includegraphics[width=\linewidth]{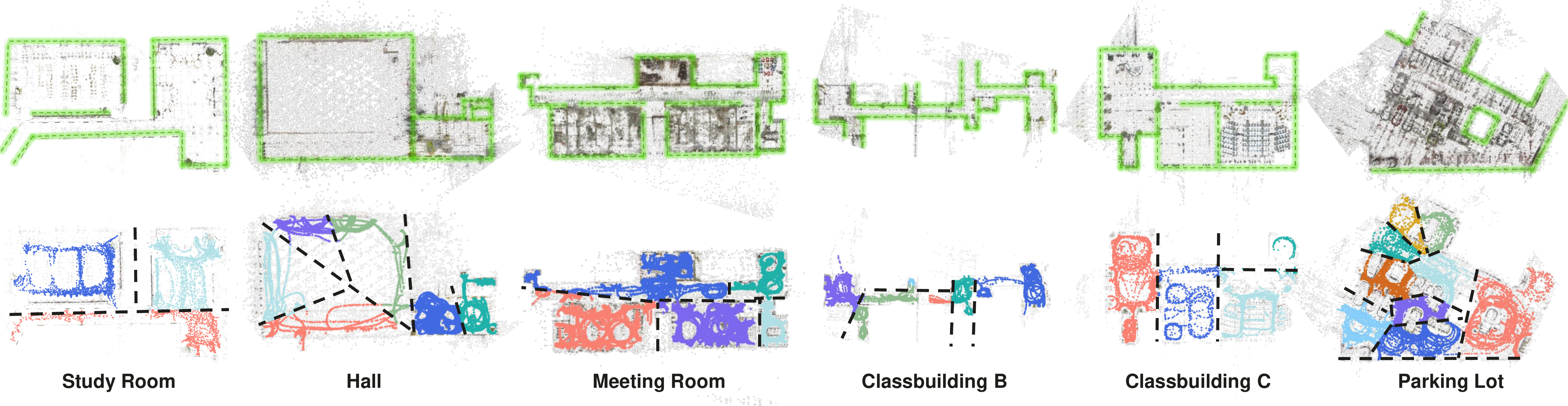}
	\caption{{Scene division on the OcclusionScene3D-E dataset.}}
	\label{fig:extended_set}
   \vspace{-2ex}
\end{figure}

\begin{figure*}[t]
	\centering
	\includegraphics[width=\linewidth]{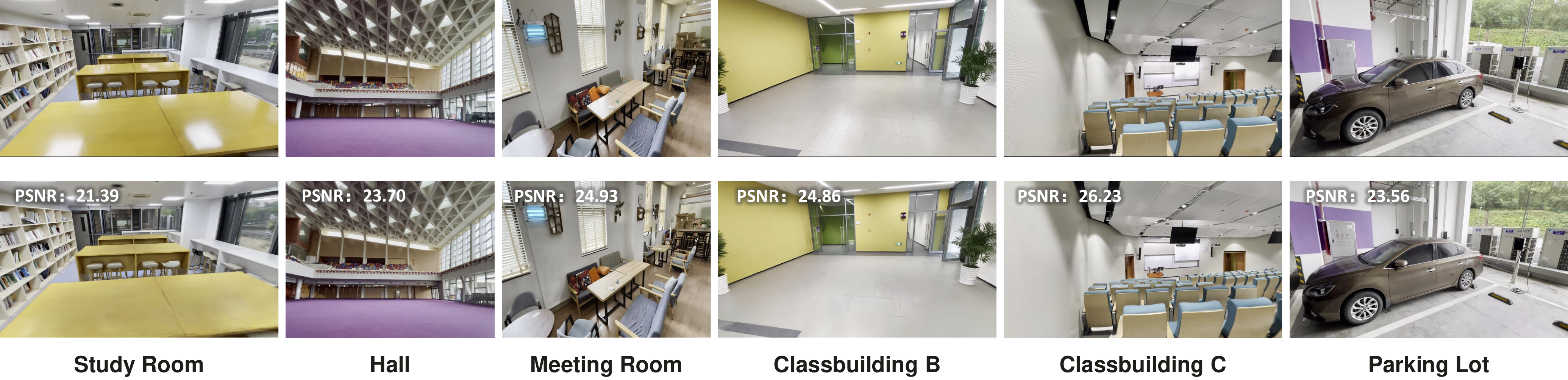}
	\caption{{Quantitative Evaluation on the OcclusionScene3D-E Dataset. The first line shows the ground truth, while the second line presents the rendering results of our OccluGaussian method.}}
	\label{fig:extended_set_rendering}
    \vspace{-1ex}
\end{figure*}

\noindent{\bf Extended camera ratios.} We further compare the extended camera ratios of division strategy with those from VastGaussian~\cite{lin2024vastgaussian}, CityGaussian~\cite{liu2024citygaussian}, Hierarchical-GS~\cite{kerbl2024hierarchical}, DOGS~\cite{chen2025dogs} and OccluGaussian, as shown in \cref{table:extended_camera_ratio}.
This ratio is defined as the total number of extended cameras across all regions divided by the total number of training cameras. A smaller ratio of extended cameras indicates that each partitioned region is more self-contained and can be well-reconstructed with fewer training iterations. Conversely, if region A and B are divided into occlusion-agnostic regions, many extended cameras for region A will be inside region B, and vice versa. In other words, these two sets of training cameras are almost the same (all cameras), making it difficult to achieve good reconstruction with limited training iterations. Compared to VastGaussian, CityGaussian, Hierarchical-GS and DOGS, our extended cameras are greatly reduced by our occlusion-aware division strategy, which leads to a higher average contribution of all training cameras for the reconstruction of each region. This advantage eventually leads to improved reconstruction quality, as proven by the results obtained when replacing our scene division strategy with others in our 3DGS optimization pipeline, as shown in \cref{table:extended_camera_ratio}.

\subsection{Additional Ablations}
\noindent{\bf Cluster number refinement.}
To obtain our final scene division, we iteratively refine the result from the graph clustering algorithm by splitting or ignoring clusters until the camera count in each cluster falls within [$M_c - \sigma_{c} M_c, M_c + \sigma_{c} M_c]$, where $M_c$ is the average camera count across all clusters. Here we explore the influence with different $\sigma_c$ choices, as shown in \cref{fig:ablation_clustering_number_strategy} and~\cref{tab:ablation_clustering_number_strategy}.
The division results remain occlusion-aware, and similar performances are obtained.
Note that without this refinement step, reconstruction quality remains comparable but demands more GPU resources for additional regions.



\noindent{\bf Features in the scene graph.} As shown in \cref{fig:clustering_features_ablation}, both the position features in the node attribute and the visibility correlations encoded by the edge weight in the scene graph are important to achieve occlusion-aware division.

\noindent{\bf Region-based rendering}. The ablation study of our region-based rendering is conducted on the OccluScene3D dataset, and the visual comparisons are shown in \cref{fig:ablation_culling_figure}. Our region-based culling strategy substantially enhances rendering speeds without any perceptible loss in visual quality. Moreover, the proposed region subdivision technique further accelerates rendering.

\noindent{\bf Training camera selection}.{The ablation study of border set cameras is shown in \cref{fig:ablation study of border set}. Without the border set, Gaussian primitives can become excessively large or elongated, leading to floaters or artifacts at region boundaries.}

\begin{table}[t]
  \centering
  \scriptsize
  \begin{tabular}{lcccc}
    \toprule
    Scene & Area ($m^2$) & \#Video & Duration ($min$) & \#Image \\
    \midrule
    \textsc{Study Room} & 500 & 39 & 105 & 4752  \\
    \textsc{Hall} & 800 & 33 & 97 & 6000  \\
    \textsc{Meeting Room} & 1500 & 36 & 127 & 10000  \\
    \textsc{ClassBuilding B} & 1000 & 27 & 63 & 7603  \\
    \textsc{ClassBuilding C} & 1000 & 25 & 76 & 6435  \\
    \textsc{Parking Lot} & 500 & 38 & 78 & 6060  \\
    \bottomrule
  \end{tabular}
    \caption{Statistics of the OccluScene3D-E dataset with six real scenes.}
  \label{table:extended_dataset_details}
\end{table}

\begin{table}[t]
\begin{center}
\resizebox{1\linewidth}{!}{
\begin{tabular}{lccccc}
\toprule
    Scene & PSNR & SSIM & LPIPS & \#GS (M) & FPS \\
    \midrule
    \textsc{Study Room} & 22.52 & 0.84 & 0.190 & 2.6 & 314.73 \\
    \textsc{Hall} & 22.70 & 0.79 & 0.212 & 1.7 & 341.76 \\
    \textsc{Meeting Room} & 27.00 & 0.90 & 0.12 & 3.5 & 407.51\\
    \textsc{ClassBuilding B} & 22.24 & 0.85 & 0.158 & 1.8 & 595.38 \\
    \textsc{ClassBuilding C} & 23.90 & 0.88 & 0.123 & 1.4 & 420.94 \\
    \textsc{Parking Lot} & 22.46 & 0.86 & 0.146 & 1.6 & 314.73 \\
    \bottomrule
\end{tabular}
}
\end{center}
\vspace{-15pt}
\caption{{Quantitative results on the OccluScene3D-E dataset.}}\label{tab:quantitative_results_extended_set}
\end{table}

\begin{table}[t]
\begin{center}
\resizebox{0.8\linewidth}{!}{
\begin{tabular}{ccccc}
\toprule
Initial $K$ & Final $K$ & PSNR & SSIM & LPIPS \\
\midrule
1 & 1 & 25.18 & 0.882 & 0.145  \\
5 & 5 & 25.35 & 0.895 & 0.114  \\
7 & 7 & 25.59 & 0.902 & 0.103  \\
10 & 7 & 25.81 & 0.903 & 0.099  \\
15 & 10 & 25.33 & 0.899 & 0.101  \\
\bottomrule
\end{tabular}
}
\end{center}
\vspace{-3ex}
\caption{Different initial clustering numbers $K$ on the Gallery scene of OccluScene3D dataset.}
\label{tab:ablation_K2}
\vspace{-3ex}
\end{table}

\section{Extended Dataset}
To further advance research in occluded scene reconstruction, we introduce an extended version of the OccluScene3D dataset named OccluScene3D-E. It encompasses six diverse scenes: \textsc{Study Room}, \textsc{Hall}, \textsc{Meeting Room}, \textsc{ClassBuilding B}, \textsc{ClassBuilding C}, and \textsc{Parking Lot}, with detailed information in \cref{table:extended_dataset_details}. As demonstrated in \cref{fig:extended_set}, applying OccluGaussian's scene division method to OccluScene3D-E yields results that closely align with the underlying scene structures. \cref{tab:quantitative_results_extended_set} details the quantitative metrics. Ablation study on initial clustering number $K$ is shown in \cref{tab:ablation_K2}. The rendering outcomes for each scene, illustrated in \cref{fig:extended_set_rendering}, collectively confirm the robustness and effectiveness of our scene division strategy, as well as its capacity to produce high-fidelity visual results. Dataset page: \url{https://occlugaussian.github.io/OccluScene3D/index.html}.

